\documentclass[twocolumn]{aastex631}

\usepackage{xspace}
\usepackage[nolist]{acronym}
\usepackage{graphicx}
\usepackage{subfigure}
\usepackage{amsmath}
\providecommand{\acrolowercase}[1]{\lowercase{#1}}

\newcommand{\pastro}{$p_\mathrm{astro}$\xspace}

\begin{acronym}
\acro{TPR}[TPR]{True Positive Rate}
\acro{FPR}[FPR]{False Positive Rate}
\acro{BNS}[BNS]{binary neutron star}
\acro{NSBH}[NSBH]{neutron star-black hole}
\acro{MDC}[MDC]{Mock Data Challenge}
\acro{EoS}[EoS]{equation of state}
\acro{NS}[NS]{neutron star}
\acro{GW}[GW]{gravitational wave}
\acro{LIGO}[LIGO]{Laser Interferometer \acs{GW} Observatory}
\acro{aLIGO}[aLIGO]{Advanced \acs{LIGO}}
\acro{AdVirgo}[AdVirgo]{Advanced Virgo}
\acro{O4}[O4]{fourth observing run}
\acro{LVK}[LVK]{LIGO-Virgo-KAGRA}
\acro{O1}[O1]{\ac{aLIGO}'s first observing run}
\acro{O2}[O2]{\ac{aLIGO}'s and \ac{AdVirgo}'s second observing run}
\acro{O3}[O3]{\ac{aLIGO}'s and \ac{AdVirgo}'s third observing run}
\acro{oLIB}[\acrolowercase{o}LIB]{Omicron+\acl{LIB}}
\acro{KAGRA}[KAGRA]{KAmioka GRAvitational\nobreakdashes-wave observatory}
\acro{GraceDB}[\texttt{GraceDB}]{GRAvitational-wave Candidate Event DataBase}
\acro{LLAI}[LLAI]{Low-Latency Alert Infrastructure}
\acro{CBC}[CBC]{compact binary coalescence}
\acro{BBH}[BBH]{binary black hole}
\acro{FAR}[FAR]{false alarm rate}
\acro{GstLAL}[GstLAL]{GStreamer LIGO Scientific Collaboration Algorithm Library}
\acro{SVD}[SVD]{singular value decomposition}
\acro{SNR}[SNR]{signal\nobreakdashes-to\nobreakdashes-noise ratio}
\acro{FGMC}[FGMC]{Farr, Gair, Mandel and Cutler formalism of Poisson Mixture Model}
\acro{MBTA}[MBTA]{Multi-Band Template Analysis}
\acro{SPIIR}[SPIIR]{Summed Parallel Infinite Impulse Response}
\acro{GPU}[GPU]{Graphical Processing Units}
\acro{cWB}[\acrolowercase{c}WB]{Coherent WaveBurst}
\acro{GCN}[GCN]{General Coordinates Network}
\acro{SCiMMA}[SCiMMA]{Scalable CyberInfrastructure for Multi-Messenger Astrophysics}
\acro{BAYESTAR}[\texttt{BAYESTAR}]{BAYESian TriAngulation and Rapid localization}
\acro{FITS}[FITS]{Flexible Image Transport System}
\acro{HEALPix}[HEALP\acrolowercase{ix}]{Hierarchical Equal Area isoLatitude Pixelization}
\acro{MOC}[MOC]{Multi-Order Coverage}
\acro{EM}[EM]{electromagnetic}
\acro{NS}[NS]{neutron star}
\acro{BH}[BH]{black hole}
\acro{LLOID}[LLOID]{Low-Latency Online Inspiral Detection}
\acro{IIR}[IIR]{infinite impulse response}
\acro{ToO}[ToO]{target of opportunity}
\acro{PE}[PE]{Parameter Estimation}
\acro{Bilby}[Bilby]{Bayesian Inference Library for \acs{CBC} \acs{GW} signal}
\acro{RAVEN}[RAVEN]{Rapid, on-source VOEvent Coincident Monitor}
\acro{AGN}[AGN]{active galactic nucleus}
\acro{GWTC}[GWTC]{Gravitational-Wave Transient Catalog}
\acro{SSM}[SSM]{Sub-Solar Mass}
\acro{MMA}[MMA]{Multi-Messenger Astronomy}
\acro{ML}[ML]{Machine Learning}
\acro{CGMI}[CGMI]{Coarse Grained Mass Information}
\acro{MEG}[MEG]{Mock Event Generator}
\acro{LLPIC}[LLPIC]{Low-Latency Production Integration Challenge}
\acro{ROC}[ROC]{Receiver Operating Characteristic}
\acro{KNN}[KNN]{K-Nearest Neighbors}
\acro{NN}[NN]{Neural Network}
\acro{MLP}[MLP]{Multilayer Perceptron}
\acro{RF}[RF]{Random Forest}
\acro{PBH}[PBH]{Primordial Black Hole}
\acro{GRB}[GRB]{Gamma-Ray Burst}
\acro{CNN}[CNN]{Convolutional Neural Networks}
\acro{ZTF}[ZTF]{the Zwicky Transient Facility}

\end{acronym}

\begin{document}

\title{Low-latency gravitational wave alert infrastructure and new data products for multi-messenger astronomy during LIGO-Virgo-KAGRA's fourth observing run}

\author[0000-0003-1314-4241]{Sushant Sharma Chaudhary*}
\affiliation{School of Physics and Astronomy, University of Minnesota, Minneapolis, Minnesota 55455, USA}
\email{*ssharmac@umn.edu}

\author[0009-0008-9546-2035]{Andrew Toivonen}
\affiliation{School of Physics and Astronomy, University of Minnesota, Minneapolis, Minnesota 55455, USA}
\affiliation{Theoretical Division, Los Alamos National Laboratory, Los Alamos, NM 87545, USA}
\collaboration{2}{these authors contributed equally to this work}

\author[0000-0003-0038-5468]{Deep Chatterjee}
\affiliation{LIGO Laboratory and Kavli Institute for Astrophysics and Space Research, Massachusetts Institute of Technology, 185 Albany Street, Cambridge, Massachusetts 02139, USA}

\author[0000-0003-1556-8304]{Roberto De Pietri} \affiliation{Dipartimento di Scienze Matematiche, Fisiche e Informatiche, Universit`a di Parma, I-43124 Parma, Italy} \affiliation{INFN, Sezione di Milano Bicocca, Gruppo Collegato di Parma, I-43124 Parma, Italy}

\author[0000-0001-6728-6523]{Alvin K. Y. Li}  
\affiliation{RESCEU, The University of Tokyo, Tokyo, 113-0033, Japan}

\author[0000-0002-8230-3309]{Cody Messick}
\affiliation{Leonard E. Parker Center for Gravitation, Cosmology, and Astrophysics, University of Wisconsin-Milwaukee, Milwaukee, WI 53201, USA}

\author[0000-0001-6331-112X]{Geoffrey Mo}
\affiliation{LIGO Laboratory and Kavli Institute for Astrophysics and Space Research, Massachusetts Institute of Technology, 185 Albany Street, Cambridge, Massachusetts 02139, USA}
\affiliation{The Observatories of the Carnegie Institution for Science, 813 Santa Barbara St., Pasadena, CA 91101, USA}
\affiliation{Department of Astronomy, California Institute of Technology, 1216 E California Blvd, Pasadena, CA 91125, USA}

\author[0000-0001-8919-0899]{Brandon Piotrzkowski}
\affiliation{Leonard E. Parker Center for Gravitation, Cosmology, and Astrophysics, University of Wisconsin-Milwaukee, Milwaukee, WI 53201, USA}

\collaboration{6}{}
\nocollaboration{38}

\author[0009-0004-2101-5428]{Shomik Adhicary} \affiliation{Department of Physics, The Pennsylvania State University, University Park, Pennsylvania 16802, USA} 
\affiliation{Institute for Gravitation and the Cosmos, The Pennsylvania State University, University Park, Pennsylvania 16802, USA}

\author[0000-0001-6589-8673]{Nicolas Arnaud} 
\affiliation{Universit\'e Claude Bernard Lyon 1, CNRS, IP2I Lyon / IN2P3, UMR 5822, F-69622 Villeurbanne, France}

\author[0000-0002-7686-3334]{Sarah Antier}
\affiliation{IJCLab, Univ Paris-Saclay, CNRS/IN2P3, Orsay, France}

\author{Ines Bentara}
\affiliation{Institut de Physique des 2 Infinis de Lyon (IP2I) - UMR 5822, Université de Lyon, Université Claude Bernard, CNRS, F-69622 Villeurbanne, France}

\author{Matthew Brodsky}
\affiliation{School of Physics and Astronomy, University of Minnesota, Minneapolis, Minnesota 55455, USA}

\author{Patrick Brockill}
\affiliation{Leonard E. Parker Center for Gravitation, Cosmology, and Astrophysics, University of Wisconsin-Milwaukee, Milwaukee, WI 53201, USA}

\author[0000-0003-0949-7298]{Sunil Choudhary}
\affiliation{Australian Research Council Centre of Excellence for Gravitational Wave Discovery (OzGrav), Australia}
\affiliation{Department of Physics, University of Western Australia, Crawley WA 6009, Australia}

\author[0000-0002-8262-2924]{Michael W. Coughlin}
\affiliation{School of Physics and Astronomy, University of Minnesota, Minneapolis, Minnesota 55455, USA}

\author[0000-0001-7663-0808]{Michael C. Davis}
\affiliation{School of Physics and Astronomy, University of Minnesota, Minneapolis, Minnesota 55455, USA}

\author[0000-0003-1354-7809]{Thomas Dent}
\affiliation{IGFAE, Universidade de Santiago de Compostela, E-15782 Spain}

\author[0000-0002-6134-7628]{Jennifer C. Driggers} \affiliation{LIGO Hanford Observatory, Richland, WA 99352, USA}

\author[0000-0002-3738-2431]{Marco Drago}
\affiliation{Universit`a di Roma “La Sapienza”, I-00185 Roma, Italy}
\affiliation{INFN, Sezione di Roma, I-00185 Roma, Italy}

\author[0000-0001-8196-9267]{Reed Essick}
\affiliation{Canadian Institute for Theoretical Astrophysics, University of Toronto, Toronto, ON M5S 3H8} 
\affiliation{Department of Physics, University of Toronto, Toronto, ON M5S 1A7}
\affiliation{David A. Dunlap Department of Astronomy, University of Toronto, Toronto, ON M5S 3H4}

\author{Bhooshan Uday Gadre}
\affiliation{Institute for Gravitational and Subatomic Physics (GRASP), Utrecht University, 3584 CC Utrecht, Netherlands}

\author[0000-0002-9370-8360]{Juan Garc\'ia-Bellido}
\affiliation{Instituto de F\'isica Te\'orica IFT-UAM/CSIC, Universidad Aut\'onoma de Madrid, Cantoblanco 28049 Madrid, Spain}

\author[0000-0001-9901-6253]{Shaon Ghosh}
\affiliation{Montclair State University, 1 Normal Ave. Montclair, NJ 07043}

\author[0000-0002-7489-4751]{Patrick Godwin}
\affiliation{LIGO Laboratory, California Institute of Technology, Pasadena, CA 91125, USA}

\author[0000-0002-3168-0139]{Matthew Graham}
\affiliation{Division of Physics, Mathematics and Astronomy, California Institute of Technology, Pasadena, CA 91125, USA}

\author[0000-0002-4320-4420]{Weichangfeng Guo}
\affiliation{Department of Physics, University of Western Australia, Crawley WA 6009, Australia}

\author{Chad Hanna}
\affiliation{Department of Physics, The Pennsylvania State University, University Park, Pennsylvania 16802, USA}
\affiliation{Institute for Gravitation and the Cosmos, The Pennsylvania State University, University Park, Pennsylvania 16802, USA}
\affiliation{Department of Astronomy and Astrophysics, The Pennsylvania State University, University Park, PA 16802, USA}
\affiliation{Institute for Computational and Data Sciences, The Pennsylvania State University, University Park, PA 16802, USA}

\author[0009-0003-6181-4526]{Theophile Jegou Du Laz}
\affiliation{Division of Physics, Mathematics and Astronomy, California Institute of Technology, Pasadena, CA 91125, USA}

\author[0000-0002-4148-4932]{Prathamesh Joshi}
\affiliation{School of Physics, Georgia Institute of Technology, Atlanta, GA 30332, USA}

\author{Erik Katsavounidis}
\affiliation{LIGO Laboratory and Kavli Institute for Astrophysics and Space Research, Massachusetts Institute of Technology, 185 Albany Street, Cambridge, Massachusetts 02139, USA}

\author[0000-0003-0710-4331]{Sergey Klimenko}
\affiliation{University of Florida, Gainesville, FL 32611, USA}

\author[0009-0000-4183-7876]{Ethan Marx}
\affiliation{LIGO Laboratory and Kavli Institute for Astrophysics and Space Research, Massachusetts Institute of Technology, 185 Albany Street, Cambridge, Massachusetts 02139, USA}

\author[0000-0003-1470-532X]{Wanting Niu}
\affiliation{Department of Physics, The Pennsylvania State University, University Park, Pennsylvania 16802, USA}
\affiliation{Institute for Gravitation and the Cosmos, The Pennsylvania State University, University Park, Pennsylvania 16802, USA}

\author[0009-0002-4724-7118]{Kira Nolan}
\affiliation{Division of Physics, Mathematics and Astronomy, California Institute of Technology, Pasadena, CA 91125, USA}
\affiliation{Department of Physics and Astronomy, University of North Carolina at Chapel Hill, Chapel Hill, NC 27599-3255, USA}

\author[0009-0003-4044-0334]{Alexander Pace}
\affiliation{Department of Physics, Pennsylvania State University, University Park, PA 16802, USA}

\author{Florent Robinet}
\affiliation{Universit\'e Paris-Saclay, CNRS/IN2P3, IJCLab, 91405 Orsay, France}

\author[0000-0002-5861-3024]{Shio Sakon}
\affiliation{Department of Physics, The Pennsylvania State University, University Park, Pennsylvania 16802, USA}
\affiliation{Institute for Gravitation and the Cosmos, The Pennsylvania State University, University Park, Pennsylvania 16802, USA}

\author[0000-0001-5726-7150]{Takahiro Sawada} \affiliation{Institute for Cosmic Ray Research, KAGRA Observatory, The University of Tokyo, 238 Higashi-Mozumi, Kamioka-cho, Hida City, Gifu 506-1205, Japan}

\author{Leo P. Singer}
\affiliation{Astroparticle Physics Laboratory, NASA Goddard Space Flight Center, Greenbelt, MD 20771, USA}
\affiliation{Joint Space-Science Institute, University of Maryland, College Park, MD 20742, USA}

\author[0000-0001-9675-4584]{Divya Singh}
\affiliation{Department of Physics, University of California, Berkeley, CA 94720, USA}
\affiliation{Institute for Nuclear Theory, University of Washington, Seattle, WA 98195, USA}

\author[0009-0003-2977-4034]{Shauryya Nirwal}
\affiliation{School of Physics and Astronomy, University of Minnesota, Minneapolis, Minnesota 55455, USA}

\author{Oscar Kane Stewart}
\affiliation{School of Physics and Astronomy, University of Minnesota, Minneapolis, Minnesota 55455, USA}

\author[0000-0002-6167-6149]{Marek J. Szczepa\'nczyk}
\affiliation{Faculty of Physics, University of Warsaw, Ludwika Pasteura 5, 02-093 Warsaw, Poland}

\author[0000-0003-2053-5582]{Colm Talbot}

\affiliation{Department of Physics, Princeton University, Princeton, NJ 08544, USA}
\affiliation{Research Computing, Princeton University, Princeton, NJ 08544, USA}

\author[0000-0002-2728-9508]{Max Trevor}
\affiliation{University of Maryland, College Park, MD 20742, USA}

\newcommand{\msun}{\ensuremath{M_{\odot}}}
\newcommand{\mc}{\ensuremath{\mathcal{M}_c}}
\newcommand{\hubbleunit}{\ensuremath{\text{km s}^{-1}~\text{Mpc}^{-1}}}
\newcommand{\totalinjections}{\ensuremath{5\times 10^4}}
\newcommand{\BNSrec}{\ensuremath{1489}}
\newcommand{\NSBHrec}{\ensuremath{1105}}
\newcommand{\BBHrec}{\ensuremath{1920}}

\newcommand{\BNSinj}{\ensuremath{40.9\%}}
\newcommand{\NSBHinj}{\ensuremath{35.8\%}}
\newcommand{\BBHinj}{\ensuremath{23.3\%}}

\newcommand{\hasns}{\ensuremath{\texttt{HasNS}}}
\newcommand{\hasremnant}{\ensuremath{\texttt{HasRemnant}}}
\newcommand{\hasgap}{\ensuremath{\texttt{HasMassGap}}}
\newcommand{\hasssm}{\ensuremath{\texttt{HasSSM}}}

\newcommand{\tsuperevent}{t_{\text{superevent}}}
\newcommand{\tmerger}{t_{\text{0}}}
\newcommand{\tevent}{t_{\text{event}}}
\newcommand{\tadvreq}{t_{\text{ADV\_REQ}}}
\newcommand{\tgcn}{t_{\text{GCN\_PRELIM}}}
\newcommand{\temcoinc}{t_{\text{EM\_COINC}}}
\newcommand{\travenalert}{t_{\text{RAVEN\_ALERT}}}
\newcommand{\gwcelery}{\texttt{GWCelery}}
\newcommand{\celery}{\texttt{Celery}}
\newcommand{\flask}{\texttt{Flask}}
\newcommand{\redis}{\texttt{Redis}}

\newcommand{\publicalertthresholdbefortrials}{\ensuremath{ \leq 1.6 \times 10^{-4}}\,Hz (fourteen per day)}
\newcommand{\publicalertthresholdaftertrails}{\ensuremath{ \leq 2.3 \times 10^{-5}}\,Hz (two per day)}
\newcommand{\significantalertthresholdbeforetrials}{\ensuremath{\leq 3.9 \times 10^{-7}}\,Hz (one per month)}
\newcommand{\significantalertaftertrialsmdc}{\ensuremath{\leq 6.4 \times 10^{-8}}\,Hz (one per 6 months)}
\newcommand{\significantalertthresholdaftertrials}{\ensuremath{\leq 7.7 \times 10^{-8}}\,Hz (one per 5 months)}
\newcommand{\burstalertthresholdbeforetrials}{\ensuremath{\leq 3.2 \times 10^{-8}}\,Hz (one per year)}
\newcommand{\burstalertthresholdaftertrials}{\ensuremath{\leq 7.9 \times 10^{-9}}\,Hz (one per 4 year)}

\newcommand{\dc}[1]{{\textcolor{red}{{[DC: {#1}]}} }}

\begin{abstract}
Multi-messenger investigations of compact object mergers, such as binary neutron star and neutron star-black hole systems, remain among the most dynamic frontiers of modern astronomy. During the fourth observing run of Advanced LIGO, Advanced Virgo, and KAGRA, the search for joint electromagnetic and neutrino counterparts to gravitational-wave signals continued using public, semi-automated data products distributed in near real-time. The sky localizations and preliminary source characterizations provided by these alerts helped guide multiple rapid follow-up observations to candidate gravitational-wave events. During this run, the low-latency search parameter space was expanded to include the sub-solar mass regime and associated public data products were developed. Further, chirp mass estimates of the binary merger system were introduced as a new public data product and distributed to enable follow-up searches to concentrate their resources on the most promising transients.  In order to streamline the implementation of new searches and data products, a mock data challenge, known as the Low-Latency Production Integration Challenge (LLPIC), was used as a replay of a simulated population of gravitational-wave sources to optimize and profile the software infrastructure and scientific deliverables.  In this paper we describe these new public data products, their performance and the result of end-to-end testing of the low-latency infrastructure with the LLPIC. 

\end{abstract}




\section{Introduction} \label{sec:intro}

The field of \ac{GW} astronomy has advanced rapidly since the landmark detection of the first \ac{BBH} merger during \ac{O1} \citep{AbEA2016b}. This was followed by the discovery of the first \ac{BNS} merger in \ac{O2} \citep{AbEA2017b}, together with its associated electromagnetic counterparts AT2017gfo \citep{CoFo2017,SmCh2017,AbEA2017f} and GRB\,170817A~\citep{GoVe2017,SaFe2017,AbEA2017e}, and by the detection of the first two \ac{NSBH} mergers \citep{AbEA2021} in \ac{O3}. Building on these milestones, the global network of advanced ground-based interferometers entered the \ac{O4} with LIGO, Virgo, and KAGRA operating in coordination. The first part of the observing run, named \ac{O4}a, yielded 128 confident compact binary mergers \citep{gwtc4} and similarly, the second part of the observing run yielded 161 such events \citep{gwtc5}. Throughout \ac{O4}, the \ac{LVK} collaboration released a high volume of real-time gravitational-wave alerts, enabling rapid electromagnetic and neutrino follow-up crucial for advancing \ac{MMA}. These efforts support a broad range of science goals, including constraining the \ac{EoS} of a \ac{NS} \citep{BaJu2017, MaMe2017, CoDi2018b, CoDi2018, CoDi2019b, AnEe2018, MoWe2018,RaPe2018,Lai2019,DiCo2020,Huth:2021bsp}, measuring the Hubble constant \citep{CoDi2019,CoAn2020,2017Natur.551...85A,HoNa2018,DiCo2020}, and probing $r$-process nucleosynthesis \citep{ChBe2017,2017Sci...358.1556C, CoBe2017,PiDa2017,RoFe2017,SmCh2017,WaHa2019,KaKa2019}. 

The \ac{LVK}'s \ac{LLAI} is a composite system comprised of multiple subsystems that enable the real-time analysis of \ac{GW} candidates and the generation of alerts. The overarching infrastructure collects \ac{GW} triggers from various low-latency search pipelines \citep{Allene_mbta,Becca_gstlal, DalCanton:2020vpm, chu2021spiir, KlVe2016, Marx:2024wjt, mly}, produces annotated data products—including classifications, source properties, and sky localizations—and issues automated public alerts.  This system plays a crucial role in supporting \ac{MMA} and \ac{ToO} searches.  A detailed description of the components of the \ac{LLAI} and the \ac{LVK}'s public data products is provided in \citet{O3_mdc}. Prior to \ac{O4}, the \ac{O3} Replay \ac{MDC} exercise \citep{O3_mdc} demonstrated the performance of these data products, characterized the various latencies involved in alert processing, and served as a reference dataset for continuous development and integration testing by both \ac{LLAI} developers and search pipeline teams.

During \ac{O4}, the \ac{LLAI} was augmented with new search capabilities,  and public data products that expand the scope of low-latency gravitational-wave alerts. In particular, public alerts for \ac{SSM} candidates were enabled, extending the low-latency \ac{CBC} parameter space beyond the stellar-mass regime \citep{Allene_mbta, chad_ssm_template}. To support interpretation of these alerts, a new public data product, \texttt{HasSSM}, was introduced. This product is based on a machine-learning classifier and provides a quantitative assessment of the probability that a merger contains an \ac{SSM} object. Additionally, the \ac{LVK} introduced a chirp mass data product consisting of binned source-frame chirp mass estimates for significant \ac{GW} candidate events, enabling more direct low-latency access to intrinsic source properties. The robustness and integration of these new searches and data products were evaluated using a dedicated, week-long \ac{MDC}, the \ac{LLPIC}, generated with simulated \ac{O4} background noise. The \ac{LLPIC} included \ac{CBC} signals from \ac{BNS}, \ac{NSBH}, and \ac{BBH} systems, as well as \ac{SSM} sources. In addition, the dataset contained transient \ac{GW} signals not associated with \ac{CBC} mergers (“bursts”), representing a range of astrophysical scenarios including core-collapse supernovae~\cite{Janka:2012wk,Abdikamalov:2020jzn}, magnetars~\cite{Nazari:2020ihd,Contopoulos:2023iom,Kouvaris:2024ait}, neutron-star starquakes~\cite{Watts:2006mr,Giliberti:2021zxn}, nonlinear memory~\cite{Ebersold:2020zah}, hyperbolic compact-object encounters~\cite{Bini:2023gaj}, and cosmic strings~\cite{Damour:2000wa,Damour:2001bk,Damour:2004kw}. Compared to the \ac{O3} replay \ac{MDC}, the inclusion of these additional signal classes provided an improved testbed for ongoing \ac{LLAI} development and validation.

To facilitate replay and validation of \ac{LLPIC} events independently of real-time search pipelines, the \ac{LVK} developed the \ac{MEG},\footnote{\url{https://git.ligo.org/emfollow/mock-event-generator}}
a software package that replays previously recorded \ac{GW} triggers and their associated data products. We used \ac{MEG} extensively after the initial pipeline analyses to validate downstream processing and associated data products. A brief description of \ac{MEG} is provided in Appendix~\ref{meg}.


This paper details these significant enhancements to the \ac{LVK} low-latency infrastructure for \ac{O4}, focusing on the new data products and the validation framework that supports them. In Sec.~\ref{LLAI}, we briefly describe the \ac{LLAI} and how public alerts for \ac{GW} are generated. In Sec.~\ref{SSM_section}, we describe the extension of public alerts to include sub-solar mass candidates, detailing the training and performance of the \texttt{HasSSM} machine learning classifier designed to identify them. Following this, in Sec.~\ref{CGMI}, we introduce chirp mass estimates, outlining the methodology and performance of this new low-latency data product. In Sec.~\ref{LLPIC}, we present the results from the \ac{LLPIC} which provided a comprehensive end-to-end validation of these new features using a realistic \ac{O4} environment. We conclude in Sec.~\ref{sec: conclusion} with a summary of these changes and a future outlook.

\section{Low-Latency Alert Infrastructure} \label{LLAI}

The \ac{LVK}'s \ac{LLAI} is a complex, distributed system that enables the real-time detection, analysis, and dissemination of \ac{GW} alerts. It integrates several critical components: low-latency data calibration and transfer, online modeled and unmodeled search pipelines, and the suite of infrastructure elements for event management and alert distribution. These include \ac{GraceDB}\footnote{\url{https://gracedb.ligo.org/}}
, which serves as the central event database; \texttt{igwn-alert},\footnote{\url{https://igwn-alert.readthedocs.io}}
 an internal Kafka-based alert system for communicating event-state updates; and \gwcelery,\footnote{\url{https://git.ligo.org/emfollow/gwcelery}}
 a distributed task queue that clusters, annotates, and orchestrates events, as well as publishes public alerts for the broader astronomical community.

The process of detection of \ac{GW} and subsequent alerts begins at the detectors. The real-time \ac{GW} strain data is calibrated at the detector sites and sent to data processing sites where the majority of low-latency search pipelines operate. The low-latency search pipelines operate continuously on this \ac{LVK} data stream to identify potential \ac{GW} signals in near real-time. Modeled searches (\ac{CBC})—including \ac{GstLAL} \citep{Messick:2016aqy, Tsukada:2023edh, Becca_gstlal}, PyCBC \citep{Nitz:2018rgo,DalCanton:2020vpm}, \ac{MBTA} \citep{Allene_mbta, Aubin:2020goo}, \ac{SPIIR} \citep{chu2021spiir, kovalam_2022}, and Aframe \citep{Marx:2024wjt}—use matched filtering or \ac{ML} based techniques to detect \ac{GW} signals of a \ac{CBC}. Unmodeled searches (Burst)—such as \ac{cWB} \citep{klimenko2008coherent, KlVe2016, drago2021coherent, Mishra:2024zzs} and MLy \citep{mly}—rely on excess power or convolutional neural network (CNN) algorithms to identify generic transient signals. When a search pipeline identifies a candidate (also called a \ac{GW} trigger), it assigns a \ac{FAR} and uploads the trigger to \ac{GraceDB}. Multiple triggers may arrive nearly simultaneously. Through \texttt{igwn-alert}, \gwcelery~continuously monitors \ac{GraceDB} for updates, performs further annotation (such as rapid sky localization and source classification), and clusters temporally coincident triggers within $\pm1\,$s into a \textit{superevent}. A preferred event is then selected—based on the highest network \ac{SNR} for \ac{CBC} pipelines or the lowest \ac{FAR} for Burst pipelines. The preferred event's data products form the basis for public alert generation once a predefined \ac{FAR} threshold is crossed (see \citealt{O3_mdc} for details). Up-to-date information about current \ac{FAR} thresholds can be found in the \ac{LVK} Public Alert User Guide\footnote{\url{https://emfollow.docs.ligo.org/userguide/}}.

Each public \ac{GW} alert is distributed with annotation data products that provide additional astrophysical context. Sky localization maps (sky maps) are generated using \ac{BAYESTAR} \citep{SiPr2016} for \ac{CBC} triggers and by dedicated pipelines for Burst triggers. For \ac{CBC} events, supplementary products—including \pastro \citep{pycbc_pastro_o4, PhysRevD.91.023005} and EM-Bright \citep{ChGh2020, Berbel_2024}—are provided to inform source classification and inferred physical properties. Prior to \ac{O4}, these classifications were limited to compact objects with component masses above $1\,M_\odot$, reflecting the mass range of the low-latency \ac{CBC} template banks. The introduction of sub-solar mass (\ac{SSM}) searches in \ac{O4} therefore necessitated the development of new annotation products to characterize and communicate potential \ac{SSM} sources.

\section{Sub-Solar Mass Search and Data Products}\label{SSM_section}

Previously, all \ac{CBC} low-latency searches had limited their search to compact objects above one solar mass; however, new \ac{SSM} searches specifically intended for objects below one solar mass have been introduced in \ac{O4} \citep{O4_SSM}.  Largely, searches for sub-solar mass compact objects are motivated by the possibility to explore new physics, and potentially detect dark matter since low-mass objects more compact than white dwarfs cannot be produced by standard astrophysical channels.

The most commonly invoked candidates are primordial black holes (PBHs) which are hypothesized to have formed from large density fluctuations in the early universe, potentially during cosmic inflation \citep{hawking_BH,carr_PBH}. While the mass spectrum of PBHs may overlap with that of stellar-origin black holes, the existence of SSM PBHs remains an open question. Notably, even the first \ac{BBH} event, GW150914, was hypothesized to be of primordial origin \citep{bird_DM, Sasaki:2016jop}, adding to the long-standing debate on whether PBHs could contribute significantly to dark matter \citep{bird_DM,clesse_PBH}.

Another proposed scenario involves the cooling and gravitational collapse of dissipative dark matter halos, which can produce BHs with sub-solar masses and lead to merger rates that may be within the sensitivity reach of advanced \ac{GW} detectors \citep{Shandera_ssm,Kouvaris_DM}.

Additionally, alternative exotic compact objects such as boson stars, composed of scalar field condensates, and novel formation channels like imploding neutron stars mediated by interactions with dark matter, could contribute to the mass spectrum near a solar mass \citep{Cardoso_review,Schunck_boson_star}. These scenarios fall within the scope of ongoing and future \ac{SSM} \ac{GW} searches.

Beyond the dark matter channels, sub-solar mass neutron stars have been proposed to form through the fragmentation of collapsar disks or core fission inside a core-collapse supernova producing a ``superkilonova"~\citep{Piro:2006ja, Metzger:2024ujc}.
Therefore, triggers from searches for coalescing \ac{SSM} binaries form interesting candidates for multi-messenger follow-up to also probe neutron star formation.

Although no confirmed \ac{GW} detections from SSM binary mergers have been reported, recent \ac{LVK} searches for sub-solar mass compact binaries in data from the second half of the third observing run have placed upper limits on their merger rate density~\citep{lvk_ssm}. These limits continue to refine our understanding of exotic compact objects and their potential role in astrophysics and cosmology.

\subsection{Low-Latency SSM Searches}

During the fourth observing run (O4), the low-latency \ac{SSM} search was carried out by the \ac{GstLAL}~\citep{Cannon:2011rj, Messick:2016aqy, Sachdev:2019vvd, Hanna:2019ezx, Cannon:2020qnf, Sakon:2022ibh, Tsukada:2023edh,chad_ssm_template} and \ac{MBTA}~\citep{LIGOScientific:2012fcp, Adams:2015ulm, Aubin:2020goo, Allene_mbta} search pipelines. Both searches target systems with at least one \ac{SSM} compact object.

In the template bank used in the low-latency search, \ac{GstLAL} restricts the component masses to $m_{1,2} \geq 0.5\, M_\odot$,  $m_{1} \leq 10\, M_\odot$ and $m_{2} \leq(1\,M_\odot)$, unlike in the offline search, where component masses extend down to $0.2 \,M_\odot$. 
The bank contains only spin-aligned templates with $|\chi_{1,2}| < 0.3$.
The template bank is constructed with a minimal match of 0.97, using the \textsc{IMRPhenomD}~\citep{Husa_imrphenomd,Khan_imrphenomd} waveform model with a cutoff on the waveform duration set to 128\,s~\citep{chad_ssm_template}.
Since the \ac{GW} signals from these low-mass systems can last for minutes in the LIGO-Virgo sensitivity band, the filtering is restricted to 45-512 Hz to alleviate memory usage.

The \ac{MBTA} low-latency \ac{SSM} search covers the same parameter space as the offline search with the component masses $m_{1,2} \geq 0.2\ M_\odot$, $m_{1} \leq 10\ M_\odot$ and $m_{2} \leq(1\,M_\odot)$.
The \ac{MBTA} bank also contains only spin-aligned templates with the dimensionless spin magnitude restricted to $-0.1  \leq \chi_{1,2} \leq 0.1$ if $ m_{1,2} \leq 0.5\,M_\odot$ while $-0.9 \leq \chi_{1,2} \leq 0.9$ for all other component masses.
The bank was constructed with a minimal match of 0.97 using the \textsc{TaylorF2} waveform approximant~\citep{taylorf2}.
\ac{MBTA} uses a low frequency cut-off of 45 Hz, and a high frequency cut-off of 1024 Hz.

For more details on the construction and performance of the template bank, and configuration of low-latency \ac{SSM} searches, see~\cite{chad_ssm_template} and~\cite{Allene_mbta}.

\subsection{SSM Injection Study}
In the $m_1-m_2$ space, chirp mass forms an ambiguity contour leading to a degeneracy where binaries with different component masses can have the same chirp mass. Due to this, the chirp mass of templates in the \ac{SSM} bank coincides with the typical chirp mass of \ac{BNS} and \ac{NSBH} signals causing \ac{SSM} searches to detect \ac{BNS} and \ac{NSBH} signals. It is therefore important to provide public data products indicating probabilities for presence of SSM, NS, and even mass-gap compact objects. To create such data products, we adopt an \ac{ML} based approach similar to \cite{ChGh2020, Berbel_2024}. 

In order to train the \ac{ML} classifier, an injection set was constructed spanning the parameter space of the online \ac{SSM} searches. Approximately $1.3 \times 10^5$ simulated CBC waveforms were used for injection with intrinsic properties (masses and spins) given in Table \ref{table:ssm_inj_dist}. The optimal network \ac{SNR} is greater than 5 for all injections and is set to prevent ``hopeless'' injections, which are unlikely to be detected
in reality. IMRPhenomXAS \citep{imrphenomxas} was used as the waveform approximant for the injections. The injection density was set to be 1 per 10\,s and the injection study was carried out in O4a data starting from Sep 06, 2023 (GPS time:1378079744) lasting two weeks. The events were distributed uniformly in co-moving volume assuming 
flat $\Lambda$CDM cosmology with $H_0 = 67.3 \, \rm{km \, s^{-1} Mpc^{-1}}$ and $\Omega_m = 0.3$ based on Planck 2018 results mentioned in Table 1 of \cite{Planck}. Both the \ac{MBTA} and \ac{GstLAL} pipelines analyzed the injection with their online template banks and produced triggers in response. Additional cuts on mass ratio ($q$) $>0.1$, detector frame chirp mass ($\mathcal{M}^{det}$) $ <=2.56$, redshift $< 0.25$ were also applied.
\begin{table*}[t]
    \caption{Distribution of intrinsic properties (component masses $m$ and spins $\chi$) of binary systems 
    in the SSM injections dataset.}
    \centering
    \begin{tabular}{ccccc}
    \multicolumn{5}{c}{\textbf{SSM Injection Properties}} \\[0.3cm] \hline
     Object & $m/M_\odot$ (min/max) & $m$ distribution & Max $\chi$ & $\chi$ distribution \\
     Primary & 0.19 - 11 & $m^{-1}$ & 0.5 & uniform \& isotropic \\
     Secondary & 0.19 - 5  & $m^{-1}$ & 0.5   & uniform \& isotropic \\ \hline
    \end{tabular}
    \label{table:ssm_inj_dist}
\end{table*}

\begin{figure}[h]
    \centering
    
    \includegraphics[width=0.5\textwidth]{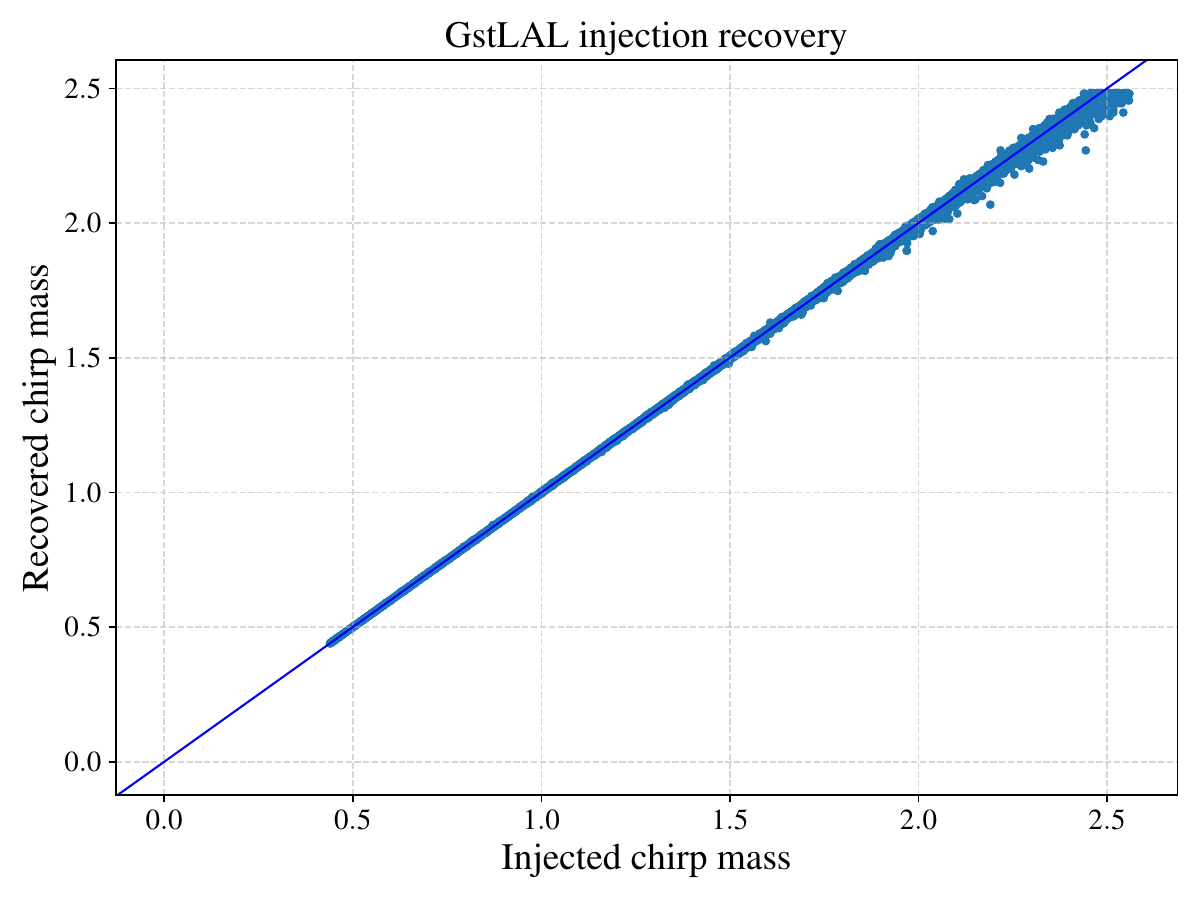}
    \includegraphics[width=0.5\textwidth]{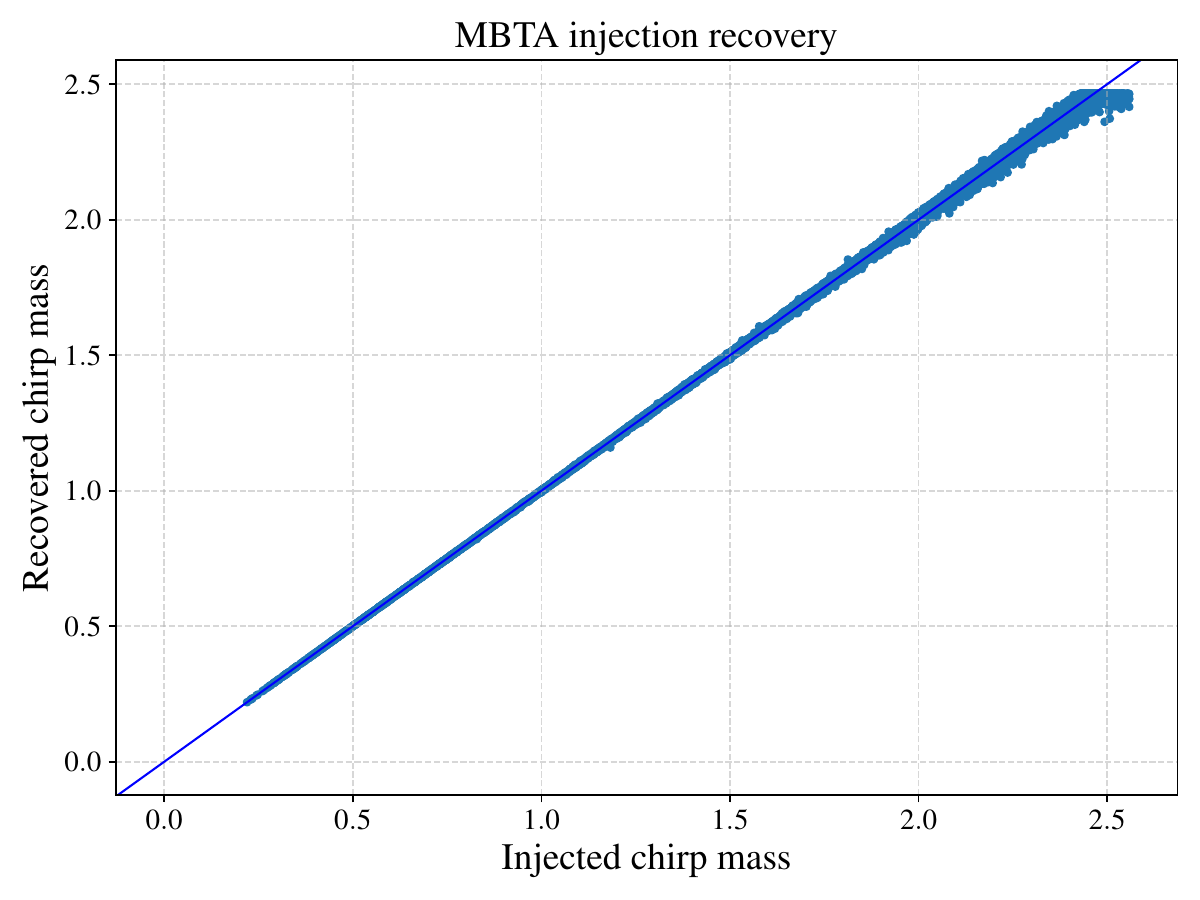}
    \caption{ Chirp mass injection-recovery plots for the GstLAL (upper) and MBTA (lower) pipeline. The diagonal line in both plots is the x=y line which indicates perfect recovery. Both pipelines accurately recover the chirp masses of the injections.}
    \label{fig:inj_rec_gstlal}
\end{figure}

Fig.~\ref{fig:inj_rec_gstlal}  shows the injection recovery from
\ac{GstLAL} and \ac{MBTA} pipelines respectively after a FAR cut of $FAR < 3.8\times 10^{-7}$ Hz. \ac{MBTA} recovered $\sim 1.3 \times 10^4$ injections whereas \ac{GstLAL} recovered $\sim 1.5 \times 10^4$ injections.
For both pipelines, chirp mass is extremely well-recovered.  
However, mass ratio $q$ is poorly recovered, typically estimating lower values than what was injected. This causes non-\ac{SSM} triggers to be recovered in \ac{SSM} parameter space as seen in Fig. \ref{fig:ssm_recovery}. This injection recovery data was used for training the classifier for individual pipelines.

\subsection{SSM Search Data Products}\label{SSM_search}
\begin{figure*}
    \centering
    \includegraphics[width=0.49\linewidth]{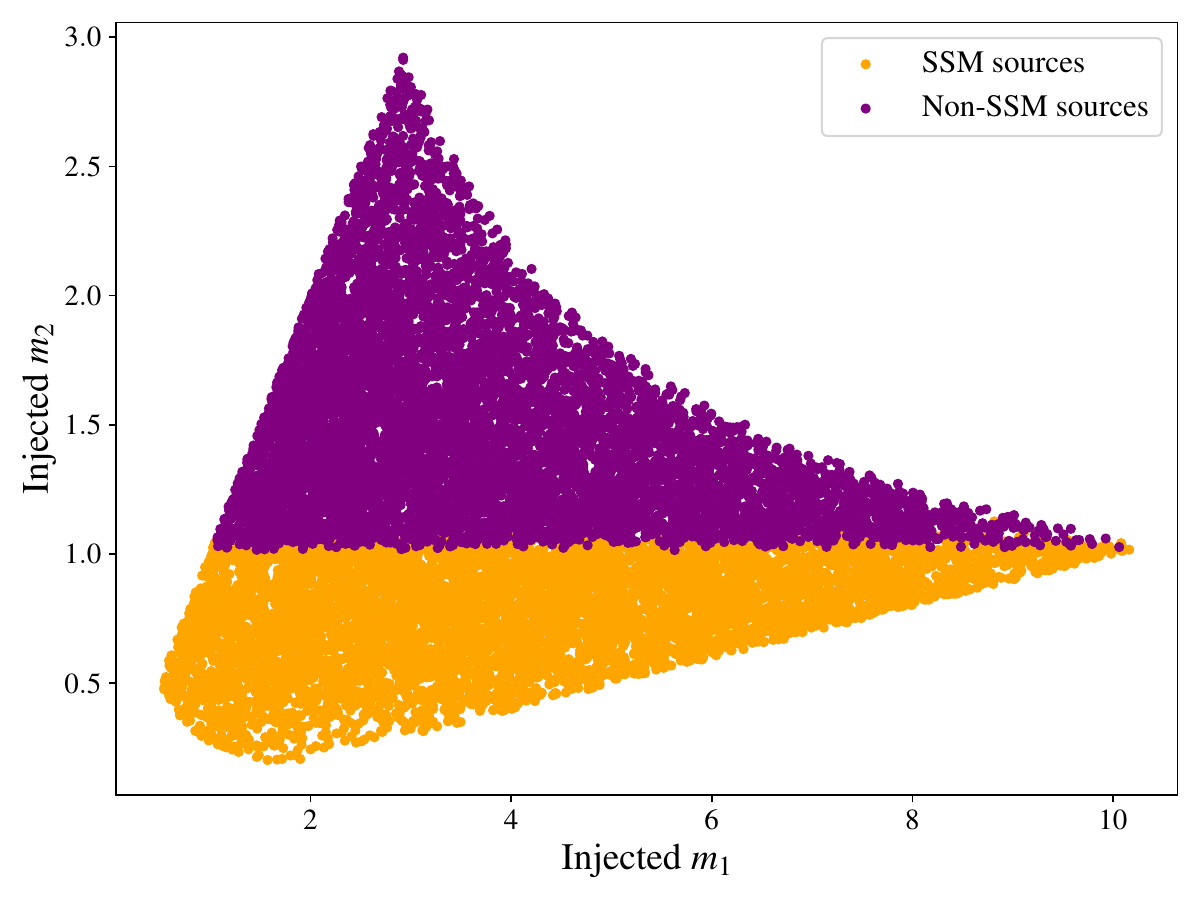}
    \includegraphics[width=0.49\linewidth]{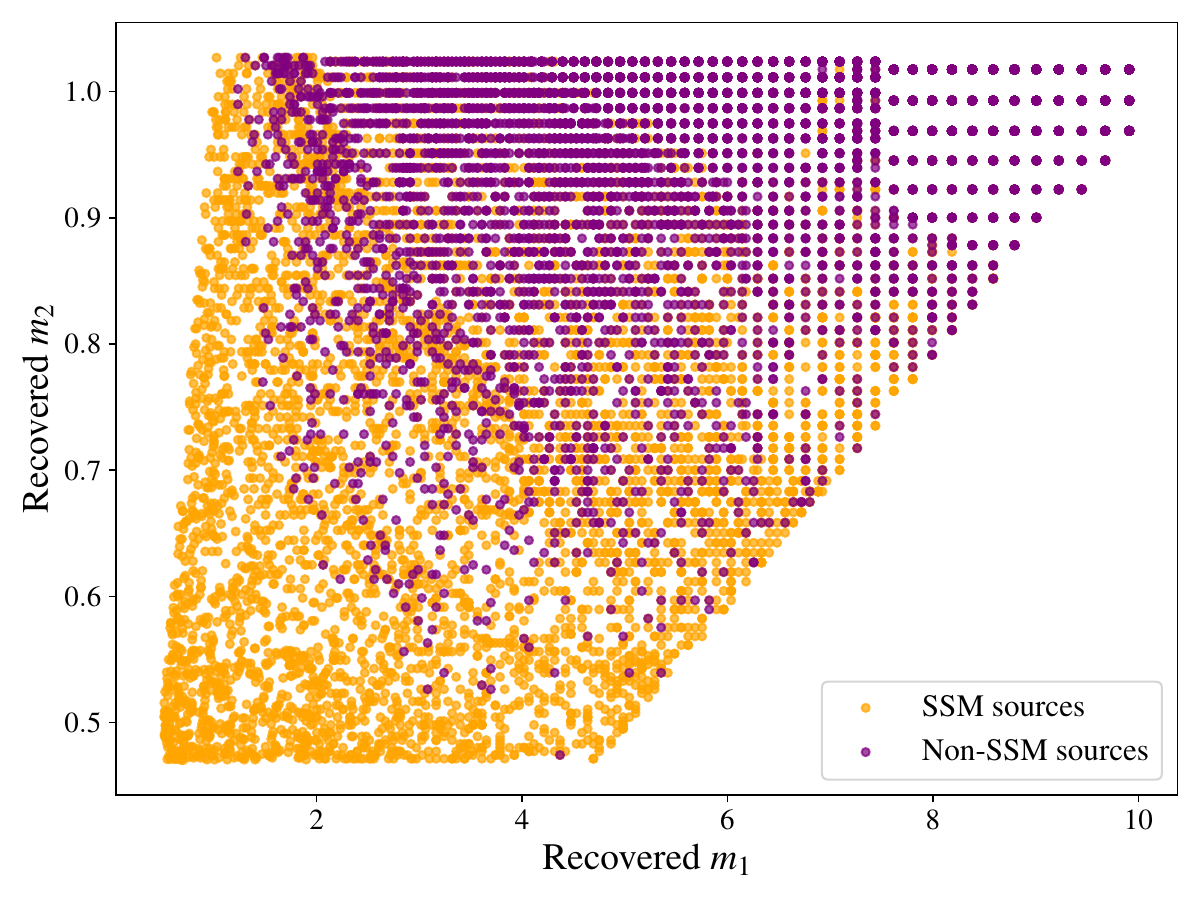}\\
    \includegraphics[width=0.49\linewidth]{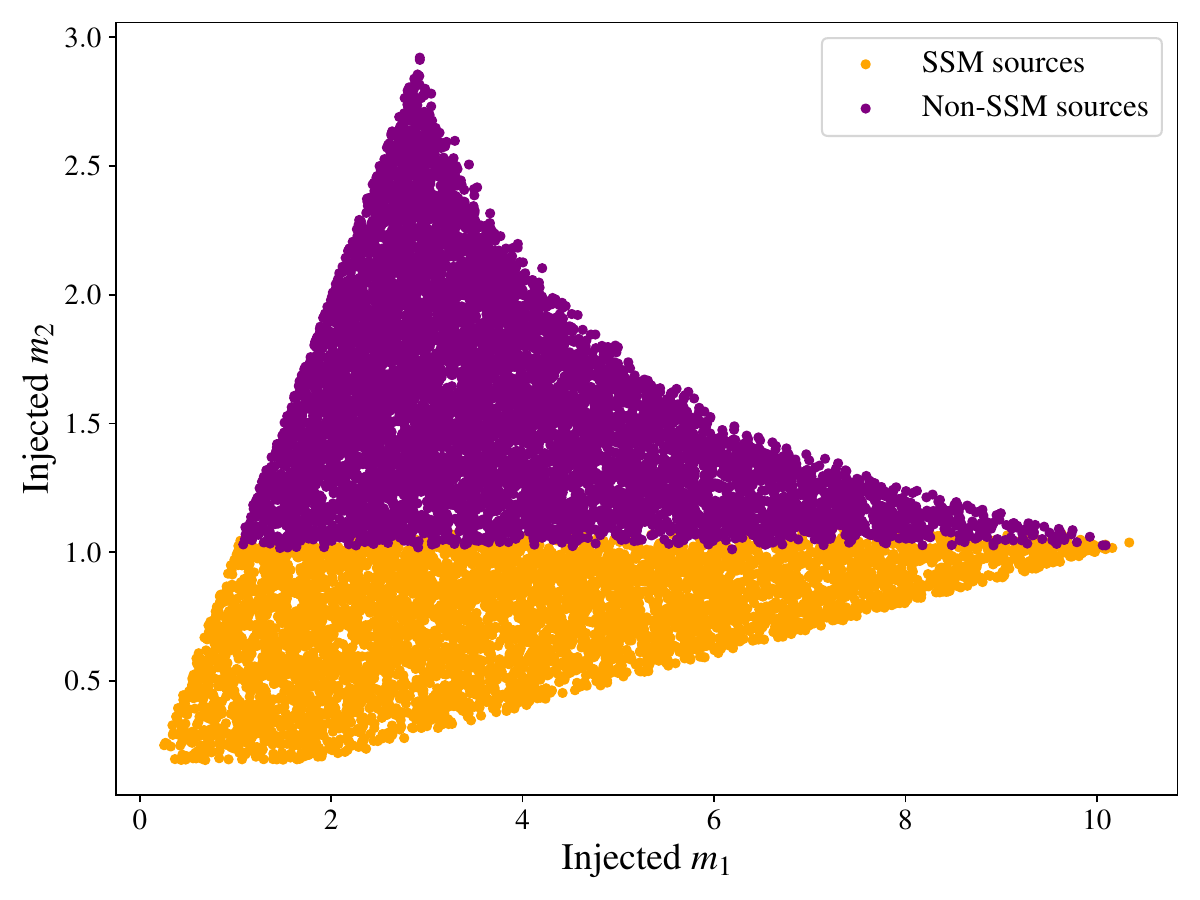}
    \includegraphics[width=0.49\linewidth]{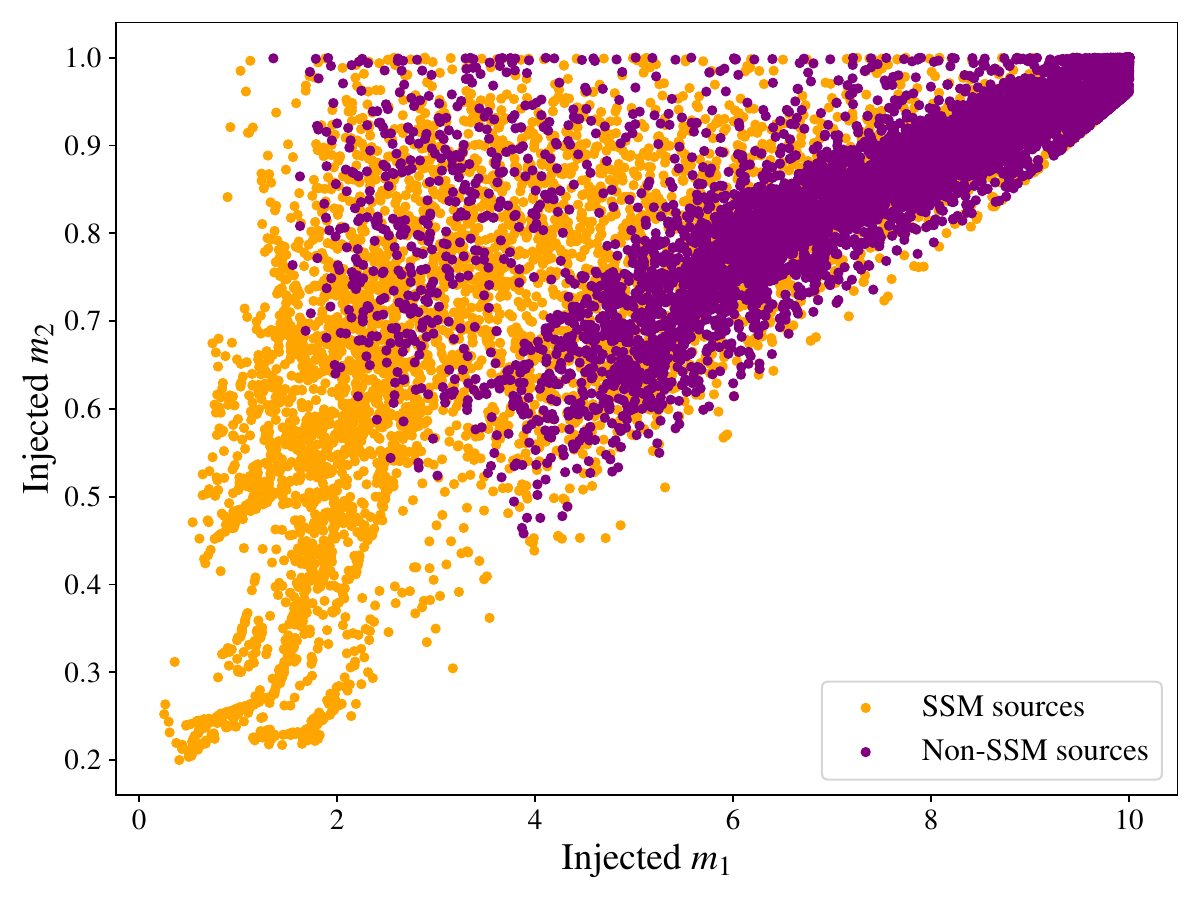}
    \caption{Scatter plots of the SSM injection dataset in the $m_1$--$m_2$ plane. \textit{Top row:} injections successfully recovered by \ac{GstLAL}, shown with injected (true) parameters (\textit{left}) and recovered (template) parameters (\textit{right}). \textit{Bottom row:} the same for \ac{MBTA}, with injected parameters (\textit{left}) and recovered parameters (\textit{right}). Colors indicate whether the injection lies in the SSM mass region or otherwise. A significant number of non-SSM injections are recovered by the SSM search with template parameters falling in the SSM region.}
    \label{fig:ssm_recovery}
\end{figure*}

As witnessed in Fig. \ref{fig:ssm_recovery}, \ac{SSM} searches can recover signals beyond the SSM parameter regime with significant FAR. The \ac{LVK} already provides the EM-Bright suite of data products \citep{ChGh2020} to characterize the source properties of \ac{CBC} candidates. During \ac{O3}, these consisted of \hasns, the probability that the binary contains a \ac{NS} component, and \hasremnant, the probability that the merger leaves remnant matter post-merger in the form of dynamical or tidal ejecta. At the start of \ac{O4}, a third product, \hasgap, was added, giving the probability that the binary contains at least one object in the lower mass-gap regime between $3-5\,M_\odot$. These quantities were, however, defined for searches restricted to component masses above $1\,M_\odot$. To provide an analogous characterization for \ac{SSM} triggers in low latency, three data products were constructed:

\begin{itemize}
    \item \texttt{HasSSM} : Probability that at least one \ac{SSM} compact object is present in the binary merger.
    \item \hasns :  Probability that at least one \ac{NS} is present in the binary merger. For SSM searches, we have defined the lower \ac{NS} boundary as $1 \,M_\odot$ and the upper \ac{NS} boundary as $3 \,M_\odot$.
    \item \hasgap : Probability that at least one mass gap compact object (3-5 $M_\odot$) is present in the
    binary merger.
\end{itemize}

The \hasns~ data product is also defined for traditional non-SSM  \ac{CBC} searches. Unlike the \ac{SSM} search, \hasns~ for other searches consider only the upper boundary of \ac{NS} based on various \ac{EoS} and the final value is marginalized over multiple \ac{EoS}. Another important quantity, \hasremnant, is not computed for \ac{SSM} searches since we do not distinguish between \ac{SSM} triggers containing a \ac{NS} and those containing a \ac{BH}, only that it has a sub-solar mass component.
 
To construct these data products, we trained on the injection dataset using a multi-labeling scheme. Each subsection of the mass parameter space (SSM-SSM, SSM-BNS, BNS-MassGap, etc.) was assigned a separate label. Although the injections were carried out with max $m_1$ and max $m_2$ of 11 $M_\odot$ and 5 $M_\odot$ respectively, the injections recovered by the pipelines were limited to $\sim$ 11 $M_\odot$ for  $m_1$ and $\sim$ 3 $M_\odot$ for  $m_2$. The dataset was further divided into training and testing sets using a 70:30 split.

Following \cite{ChGh2020} and \cite{Berbel_2024}, both \ac{KNN} and \ac{RF} algorithms were used in the training process. The input data for training consisted of reconstructed 
parameters \{$m_1^{rec}$, $m_2^{rec}$, $\chi_{1,z}^{rec}$, $\chi_{2,z}^{rec}$, $\mathcal{M}^{rec}$, SNR\} as input features. After some hyperparameter tuning and  cross-validation, the \ac{KNN} algorithm with distance-based weights and 25 neighbors was chosen for the training purpose since it slightly outperformed \ac{RF}.

\begin{figure}[h]
    \centering
    
    \includegraphics[width=0.5\textwidth]{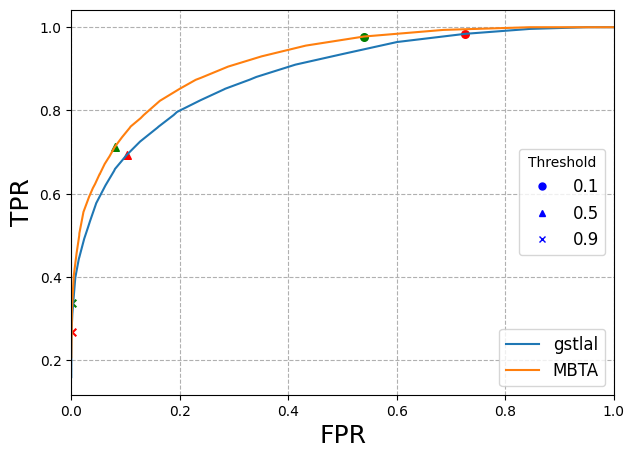}
    \includegraphics[width=0.5\textwidth]{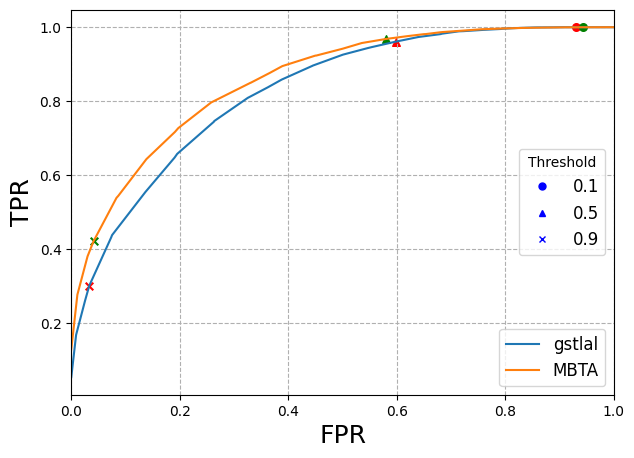}
    \includegraphics[width=0.5\textwidth]{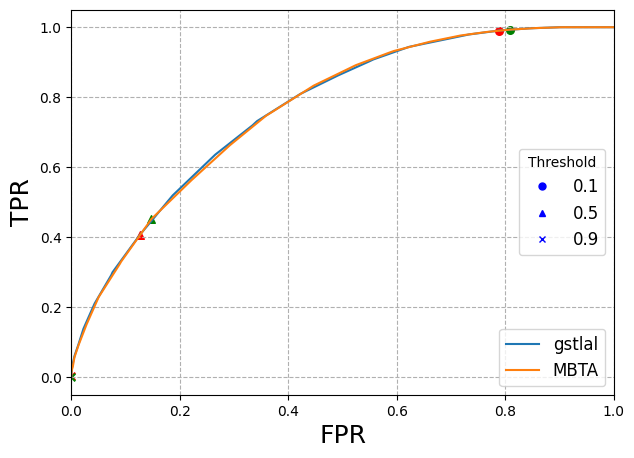}

    \caption{ROC curves on the top panel represent \texttt{HasSSM}, the middle represents  \hasns~ and the bottom panel represents \hasgap. The two different lines represent classifiers trained for individual pipelines. The markers represent the threshold used within the ROC curve.}
    \label{roc_ssm}
\end{figure}

The final data products \hasssm, \hasns, and \hasgap~were computed by combining probabilities of  labels contained within the definition of each data product. Using the testing set, \ac{ROC} curves were generated showing the performance for each data product in Fig.~\ref{roc_ssm}. The \ac{ROC} curves show that
\texttt{HasSSM}~ performs better compared to other data products. The SSM searches primarily target sources in the region $m_2 < 1 M_\odot$ and the classifier shows $\sim$ 0.85 \ac{TPR} at 0.2 \ac{FPR}. MBTA shows better performance than GstLAL, which can be linked to the greater coverage of the MBTA bank with the SSM injections used to train the classifier, in both masses and spins. \ac{SSM} searches are not designed to pick signals in the \ac{BNS} and \ac{NSBH} regimes. Despite that, they can recover \ac{BNS} and \ac{NSBH} with significant \ac{FAR} as seen in the injection-recovery plots. Signals in this region are typically identified with lower $q$ values, and it becomes challenging for the classifier to learn this behavior consistently. This can be reflected in the poorer performance of \hasns~ and
\hasgap~ (see Fig.~\ref{roc_ssm}).    

\subsection{Low-Latency Implementation}
The low-latency \ac{SSM} pipeline data products have been publicly available since early 2025. In the current implementation, a significant \ac{GW} trigger from \ac{SSM} searches is released publicly only when no significant counterpart is identified by the standard (non-SSM) \ac{CBC} searches. In such cases, the public release includes the sky localization map along with the \ac{SSM} data products—\hasssm, \hasns, and \hasgap—contained within the \texttt{em-bright} packet. As of \ac{O4}, $p_{\mathrm{astro}}$ values are not available for \ac{SSM} searches. The significance threshold for \ac{SSM} triggers is set to a \ac{FAR} of one per year ($3.17 \times 10^{-8}~\mathrm{Hz}$) before accounting for trials factors.

Following completion of the low-latency \ac{PE}, the data products are updated based on the posterior samples. Since the introduction of \ac{SSM} searches and associated data products, the \texttt{em-bright} \ac{PE} update now includes four quantities: \hasns, \hasremnant, \hasgap, and \hasssm. The \hasssm~ quantity represents samples with component masses $< 1~M_\odot$, while \hasns~ includes samples $\geq 1~M_\odot$ constrained by the neutron star \ac{EoS}. The \hasns~ value is marginalized over 10,000 \ac{EoS} draws \citep{Legred:2021hdx, legred_isaac_2022_6502467}. The \hasremnant~ quantity is evaluated using samples from the \ac{BNS} and \ac{NSBH} regions following \cite{FoHi2018} and marginalized over a large ensemble of \ac{EoS} realizations. Finally, \hasgap~ corresponds to component masses between $3$ and $5~M_\odot$. The same set of updated data products is provided for all classes of \ac{CBC} triggers.



\section{Public Chirp Mass Estimates}\label{CGMI}


The \ac{LVK} is now providing low-latency, public mass estimates for significant candidate compact merger events. Low-latency \ac{GW} searches have traditionally provided estimates of \ac{FAR}, sky localization, and source classification, but no estimate of mass. Mass information can be used to predict the properties of potential \ac{EM} counterparts \citep[see e.g.][]{Margalit:2019dpi}. These new mass estimates are released as probabilities for a predetermined set of chirp mass bins. The chirp mass is estimated initially via a combination of search pipeline outputs and sky maps, and later updated using parameter estimation results. This data product enables better interpretation of \ac{GW} candidate events and aids multi-messenger \ac{ToO} searches by providing mass information alongside source classification to better characterize the progenitor system.




Chirp mass, $\mathcal{M}$, is defined as:

\begin{equation} \label{chirp_mass}
    \mathcal{M} = \frac{(m_1 m_2)^{3/5}}{(m_1+m_2)^{1/5}}. 
\end{equation}

A given value of $\mathcal{M}$ can correspond to a range of different individual component masses. Chirp mass is well-measured from a \ac{GW} signal due to its relationship with the frequency evolution of the signal, making \ac{GW} searches well-positioned to provide low-latency chirp mass estimates. Throughout, these estimates assume the candidate is astrophysical in origin.


A new public data product released towards the end of \ac{O4} consisted of binned estimates of chirp mass for significant events uploaded by \ac{CBC} searches, including cWB BBH \citep{klimenko2008coherent, KlVe2016, drago2021coherent}. In solar masses, the chirp mass bin boundaries are:

\begin{equation}
\begin{aligned}
   [0.1, 0.87, 1.0, 1.1, 1.2, 1.3, 1.4, 1.5, 1.7, \\ 1.9, 2.1, 2.3, 3, 5.5, 11, 22, 44, 88, 1000].
\end{aligned}
\end{equation}


These bins are in the source frame of the merger, \textit{except} in the case of the low-latency cWB BBH estimate, where there is no distance or redshift information available. In this case, we report the ``detector frame'', or redshifted estimate while still using the same bins. Conversely, in cases where we have an estimate of the source distance or redshift, we relate the redshifted detector frame value $\mathcal{M}_z$ to $\mathcal{M}$ using the redshift $z$:

\begin{equation} \label{chirp_detector_frame}
    \mathcal{M}_z = (1+z) \mathcal{M} .
\end{equation}

\begin{figure*}
\centering
    \includegraphics[width=\textwidth]{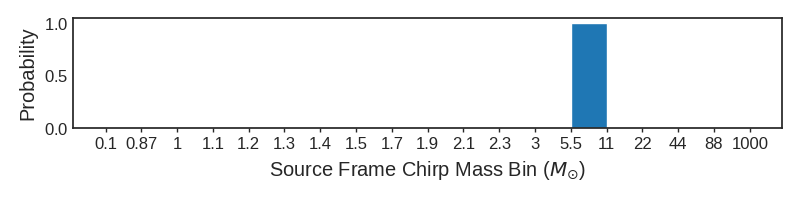}
    \caption{Example histogram of chirp mass estimate available on \ac{GraceDB} for significant events.}
    \label{fig:LL_hist}
\end{figure*}

The resulting chirp mass estimates are made publicly available on \ac{GraceDB} alongside the Second Preliminary Alert, and may later be the subject of an Update Alert. On \ac{GraceDB}, they can be found in the form of a \texttt{.json} file of the probabilities for each bin and a \texttt{.png} plot of the same probabilities, as shown in the example in Fig.~\ref{fig:LL_hist}. The highest probability bin will be reported in the circular, but
this information is \textit{not currently} included in the alert packet \texttt{.xml}. This was done to ease integration for O4; inclusion is a possible goal for O5.

\begin{figure}
\centering
    \includegraphics[width=\columnwidth]{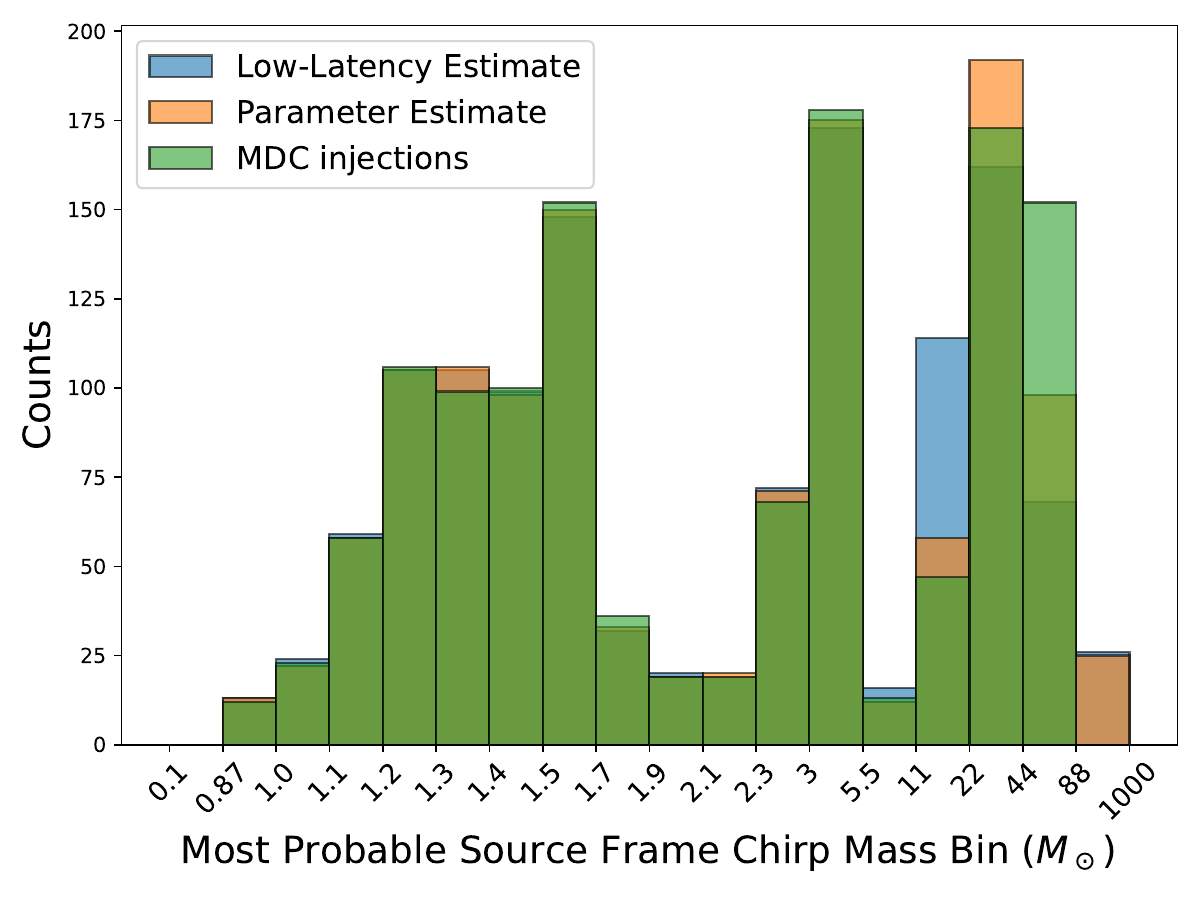}
    \caption{The most likely bin from the low-latency and parameter estimation-based estimates run on \ac{MDC} \cite{O3_mdc} events compared to their injected chirp mass.}
    \label{fig:chirp_hist}
\end{figure}

Fig.~\ref{fig:chirp_hist} shows the low-latency and parameter estimation-based estimates run on \ac{MDC} \cite{O3_mdc} events compared to their injected chirp mass values. The distribution of estimated chirp masses closely matches the distribution of injected values in bins with low mass, where multi-messenger follow-up is strongly motivated, indicating the high accuracy of the estimation in this regime.

The binned chirp mass estimates are intended to help inform time-sensitive follow-up activities. This includes multi-messenger follow-up of events with at least one neutron star \citep{Paterson:2020mmd, GRANDMA:2022fgk, Paek:2023xdc, Ahumada:2024qpr, J-GEM:2025qzq, Coughlin:2019xfb}, as well as follow-up of high mass BBH events \citep{HESS:2021gbx, Vajpeyi:2021ccn} in the search for electromagnetic counterparts that could result from the flaring and disruption of AGN accretion disks during merger. 

\subsection{Bin Rationale}

The bins were designed to be finely spaced at lower masses where chirp mass is recovered very accurately, and as these low masses are of particular interest due to their potential for  containing a \ac{NS}. While specific bins do not exclusively relate to a given source class, they were designed with the different regimes or source classes in mind, and can be loosely split into the following three regimes: the \ac{SSM} regime, the \texttt{HasNS} regime, and the \ac{BBH} regime. As we cover the different regimes, keep in mind there are multiple component mass pairs for a given  chirp mass, and that the regimes themselves are simply meant to be illustrative of the different source classes we can expect to observe.

The \ac{SSM} regime corresponds to the bin [0.1, 0.87] $M_\odot$, likely to contain an event with a component mass $< 1 M_\odot$. The upper edge of this bin, a chirp mass of 0.87 $M_\odot$, may correspond to a 1 $M_\odot$, 1 $M_\odot$ merger, for example, demonstrating that events with $\mathcal{M} < 0.87 M_\odot$ will contain a component mass $< 1 M_\odot$.
The \texttt{HasNS} regime roughly corresponds to chirp mass values for mergers with a strong possibility of containing a \ac{NS}: either \ac{BNS} or \ac{NSBH}. The bins in the chirp mass range [0.87, 3] $M_\odot$ roughly fall within this regime, again with the reminder that these regimes are not exclusive and are for illustrative purposes. For a \texttt{HasNS} example, a chirp mass of 3 $M_\odot$ may correspond to a 1.4 $M_\odot$, 10 $M_\odot$ \ac{NSBH} merger.
Lastly, the coarse, high mass bins are likely to correspond to \ac{BBH} events.  Chirp mass bins above 3 $M_\odot$, which go all the way to 1000 $M_\odot$, have a strong chance of being a \ac{BBH}. 




\subsection{Implementation in alert infrastructure}


Binned chirp mass estimates come in two forms: (i) estimates based on information available in low latency (pipeline uploads, sky maps), and (ii) estimates based on \texttt{Bilby} parameter estimation \citep{2019ApJS..241...27A, Romero-Shaw:2020owr}. The former are immediately available in low latency, while the latter accompany update alerts approximately one or more hours later when \texttt{Bilby} parameter estimation is complete. Similar to the procedure with other update information (EM-Bright, sky maps, etc), the updated chirp mass based on parameter estimation will be uploaded and identified as resulting from parameter estimation, without deleting or overwriting the original low-latency estimate.

The low-latency chirp mass estimates use the pipeline point estimate of the detector chirp mass, combined with distance information provided by the \texttt{BAYESTAR} sky map \citep{SiPr2016}, in order to correct the detector chirp mass value to the source frame. The result is binned and converted to an estimate of the probability corresponding to each bin. In the case of cWB BBH, there is no distance information, and therefore we assume a redshift of zero when reporting the source chirp mass. The estimate is uploaded in the form of a  file named \texttt{mchirp\_source.json}. 

Once the low-latency \ac{PE} is completed, the chirp mass estimates are updated using the source frame posterior samples directly. The result is binned and converted to an estimate of the probability corresponding to each bin, and that estimate will be uploaded in the form of a \texttt{.json} file named \texttt{mchirp\_source\_PE.json}.












\subsection{Multi-Messenger Applications}

In the following, we highlight two direct multi-messenger applications of the chirp mass estimates: i) informing follow-up of potential \ac{NS} mergers by helping determine whether the merger likely contained a \ac{NS}, and how likely that event is to produce a kilonova, and ii) conducting targeted follow-up of high mass \ac{BBH} events in the hope of detecting an \ac{EM} counterpart in the form of a flare from an \ac{AGN}.

\subsubsection{Neutron Star Merger Follow-up}

\ac{GW} candidate events with at least one \ac{NS} are often subjects of \ac{ToO} searches for \ac{EM} counterparts such as \acp{GRB} or kilonovae. \ac{ToO} searches rely on low-latency data products and timely decision making when determining which \ac{GW} candidate events are worth devoting telescope resources to. When searching for kilonovae, one main criterion astronomers consider is whether the candidate event likely contains a \ac{NS}. Alongside quantities such as \texttt{HasNS} and \pastro, we expect the chirp mass estimates to help answer this question. In addition, not only do these quantities offer insight on the possible presence of a \ac{NS}, they also provide an actual mass estimate. This is useful as we expect different masses to produce varying amounts of mass ejecta and consequently have varying likelihoods of producing a kilonova. The chirp mass could also be used to predict the resulting remnant of the merger \cite{Margalit:2019dpi}.

\subsubsection{\ac{BBH} Follow-up}

It is expected that high-mass \ac{BBH} mergers that take place in the accretion disks of \ac{AGN} may produce observable \ac{EM} flares \cite{Graham:2022xxu}. \texttt{BBHBot} is an automatic triggering pipeline designed to monitor \ac{GW} triggers in real time, and schedule \ac{ToO} observations with \ac{ZTF} \cite{Bel2014} to observe promising \ac{BBH} candidates. We have integrated the chirp mass estimates into \texttt{BBHBot} for the purpose of targeted follow-up of high-mass mergers. \texttt{BBHBot} ingests public low-latency \ac{GW} data products such as the new chirp mass estimates, in addition to the sky map and \ac{FAR} in order to determine which events meet the triggering criteria. If candidate events pass cuts on mass and \ac{FAR} as well as meet our observing criteria, including cuts on the 90\% sky map area and air mass, an observing plan is automatically drafted and submitted to trigger \ac{ZTF}.

\begin{figure*}
     \centering
     \includegraphics[width=.45\linewidth]{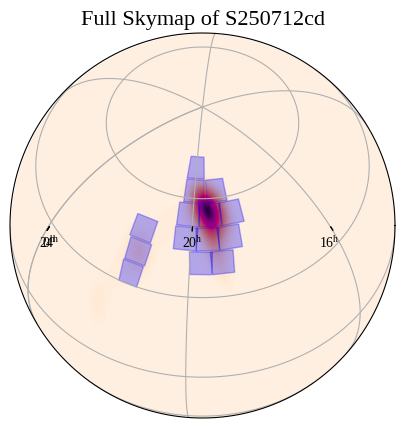}
     \centering
     \includegraphics[width=.45\linewidth]{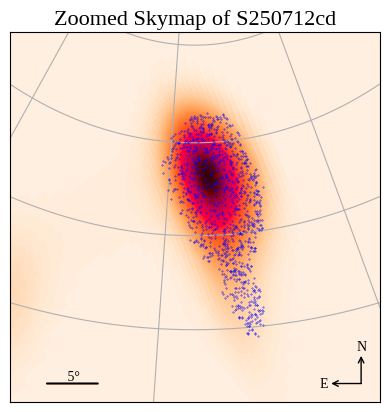}
     \caption{Targeted follow-up of the \ac{BBH} \ac{GW} candidate event S250712cd, displayed on the \ac{GW} sky localization. \ac{ZTF} telescope tilings are shown by the purple rectangles on the left, while on the right we see a zoomed version with known \ac{AGN} marked as purple dots for reference.}     
     \label{fig:bbh-bot}
\end{figure*}

Specifically, we are most interested in events with $> 100 $ $M_\odot$ of total mass, but consider those $> 50 \ M_\odot$ of total mass, for follow-up. Consequently, we filter for events most likely to fall within chirp mass bins $> 22 \ M_\odot$. Keeping in mind the definition of chirp mass in Eq. \ref{chirp_mass}, the merger of a pair of $25 \ M_\odot$ \acp{BH} corresponds to a $\mathcal{M} \sim 22 \ M_\odot$, while the merger of a pair of $50 \ M_\odot$ \acp{BH} corresponds to a $\mathcal{M} \sim 44 \ M_\odot$. Following the addition of chirp mass to the public \ac{GW} data products in June 2025, \texttt{BBHBot} has used it in the real-time evaluation of dozens of candidate \ac{BBH} mergers before the conclusion of \ac{O4}. Of these, 26 candidate events exceeded the selection threshold of chirp mass $> 22 \ M_\odot$, and 6 candidates\footnote{\href{https://gracedb.ligo.org/superevents/S250712cd/view/}{S250712cd}, \href{https://gracedb.ligo.org/superevents/S250827l/view/}{S250827l}, \href{https://gracedb.ligo.org/superevents/S251018bi/view/}{S251018bi}, \href{https://gracedb.ligo.org/superevents/S251031cq/view/}{S251031cq}, \href{https://gracedb.ligo.org/superevents/S251105aj/view/}{S251105aj}, \href{https://gracedb.ligo.org/superevents/S251108fi/view/}{S251108fi}} passed all selection criteria and were scheduled for automated observations by \ac{ZTF}. The targeted follow-up of \href{https://gracedb.ligo.org/superevents/S250712cd/view/}{S250712cd}, one such event that passed all criteria and was scheduled for observations, is seen in Fig.~\ref{fig:bbh-bot}, displayed on the \ac{GW} sky localization. On the left, the \ac{ZTF} tilings are shown by the purple rectangles, while on the right we see a zoomed version of the sky map with known \ac{AGN} marked as purple dots for reference. Those \ac{AGN} are the sources that will be observed as candidates for potential flaring events. No promising counterparts were identified in coincidence with the 6 candidates observed.


\subsection{O4 SSM Candidate Events}

\href{https://gracedb.ligo.org/superevents/S250818k/view/}{S250818k} was a recent low-significance \ac{GW} candidate of interest (with FAR of $6.810 \times 10^{-8}$ Hz) with a possible external coincidence with an \ac{EM} source\footnote{https://gcn.nasa.gov/circulars/41414}, AT2025ulz. Extensive follow-up of the counterpart AT2025ulz was carried out, with the possibility that the source was associated with the \ac{GW} event and compact object merger \citep{Franz:2025uan, Gillanders:2025fwf, Kasliwal:2025keb}. Due to this possible coincidence, chirp mass estimates were provided for this \ac{GW} event\footnote{https://gcn.nasa.gov/circulars/41437}, even though it did not cross the FAR threshold for significance. The chirp mass estimates placed S250818k in the lowest mass bin, and, assuming an astrophysical origin, the event was assigned high probabilities of containing a \ac{NS} and of containing an \ac{SSM} object. 

Later during the observing run, a second \ac{SSM} candidate, \href{https://gracedb.ligo.org/superevents/S251112cm/view/}{S251112cm}\footnote{https://gcn.nasa.gov/circulars/42650}, was detected as a significant event, unlike S250818k. S251112cm was identified with a \ac{FAR} of $5.103 \times 10^{-9}$ Hz by MBTA's dedicated \ac{SSM} search (Sec.~\ref{SSM_section}), and like S250818k, has a chirp mass estimated to lie entirely within the [0.1, 0.87] $M_\odot$ bin. Both events had \texttt{HasSSM} values of 1 based on parameter estimation, pointing to a high likelihood of containing a \ac{SSM} object, assuming the events are astrophysical in origin. Follow-up searches did not identify any credible \ac{EM} counterparts.

There has never been a confirmed \ac{SSM} detection, so this possibility is exciting and demonstrates the use of the new chirp mass data product. In addition, S251112cm is a notable example demonstrating the value of expanding searches to the \ac{SSM} regime, as this event was not found by the usual full-bandwidth \ac{CBC} searches.


\begin{figure*}
\centering
    \includegraphics[width=0.8\textwidth]{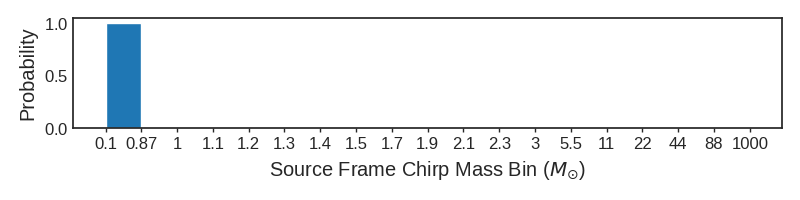}
    \caption{Histogram of chirp mass estimate for S250818k publicly available on \ac{GraceDB} and placing the candidate event in the \ac{SSM} regime.}
    \label{fig:ssm_hist}
\end{figure*}

\section{Low-Latency Production Integration
Challenge} \label{LLPIC} 

The \ac{LLPIC} was commissioned to facilitate the implementation and validation of new low-latency searches and their associated downstream data products through a dedicated \ac{MDC}. The campaign consists of a real-time replay of gravitational-wave signals injected into representative detector noise and is designed to validate search performance, assess annotation products, and ensure seamless integration with the broader \ac{LLAI}.

The background data for the \ac{LLPIC} comprise a continuous one-week segment of strain data recorded by the LIGO Hanford, LIGO Livingston, and Virgo instruments during the O4b observing period, spanning April 27--May 4, 2024 (GPS times 1398297600--1398902400). KAGRA was not observing during this period and is therefore not included. Simulated \ac{CBC} and burst waveforms were injected into this dataset and streamed in a manner identical to the low-latency data distribution used in \ac{O4} operations. The data were analyzed by the low-latency \ac{CBC} and burst search pipelines, and the resulting \ac{GW} triggers were submitted to an internal instance of \ac{GraceDB}, \texttt{GraceDB-playground}.

Subsequent event clustering, annotation, and alert generation were performed by \gwcelery, preserving the full alert-generation workflow of the \ac{O4} low-latency production system. The injected one-week dataset is replayed on a weekly cadence beginning April 29, 2025, producing multiple \ac{LLPIC} cycles that enable continuous testing and validation of the evolving low-latency infrastructure.

\subsection{CBC Injections}

A total of 2,364 simulated \ac{CBC} waveforms with mass and spin distributions detailed in Table~\ref{table:inj_dist} were injected into the \ac{LLPIC} data stretch with an interval of $\sim 4$ minutes between injections. The optimal network \ac{SNR} cutoff was set to 6 to allow recovery of more injections relative to the O3 MDC. The {\tt{IMRPhenomXPHM}} waveform approximant is used to generate waveforms for injections. For injection purposes, the boundary between \ac{NS}s and \ac{BH}s was considered to be $3 M_\odot$. The binary sources in Table ~\ref{table:inj_dist} were mixed to ensure an equal number of injections belonging to each source class. While the injection rate density used in the study is artificially high and the distribution is not representative of the true population, these injections are suitable for testing the low-latency infrastructure and data products other than \pastro, which relies on a background distribution expected for normal detector operations. The injections were carried out in all three detectors--Hanford, Livingston, and Virgo (H, L and V) \citep{aLIGO, adVirgo}--and a fraction of them fall during detector downtime. The CBC injection set consists of 600 \ac{SSM}, 626 \ac{BNS}, 553 \ac{NSBH}, and 585 \ac{BBH} injections.

\begin{figure}
\centering
    \includegraphics[width=\columnwidth]{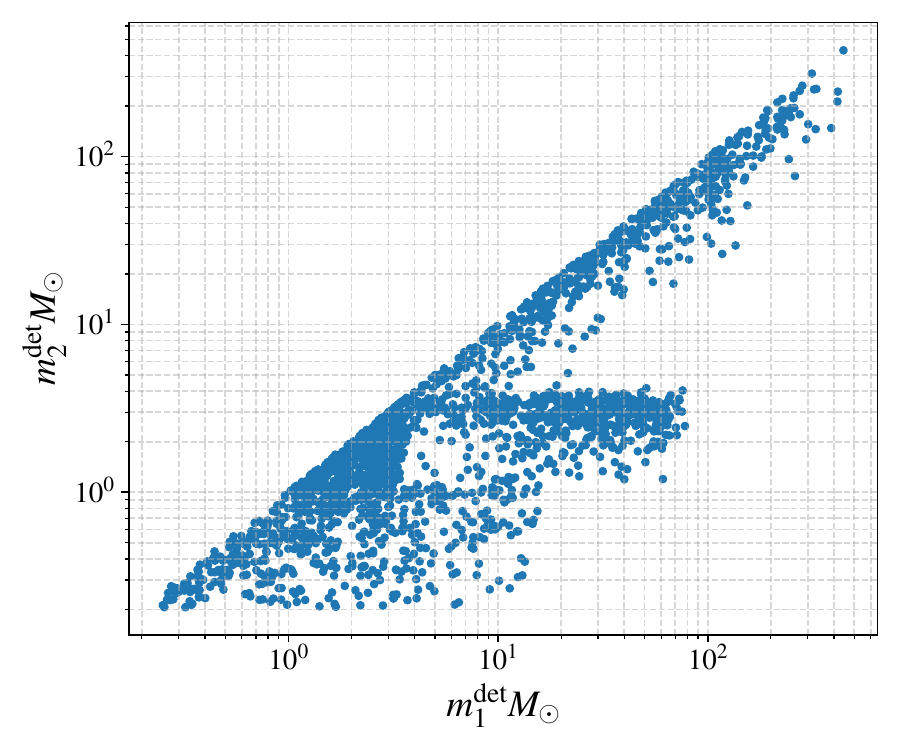}
    \caption{Distribution of LLPIC CBC injections in detector frame masses.}
    \label{fig:cbc_inj_dist}
\end{figure}

\begin{deluxetable*}{cccccc}
\tabletypesize{\footnotesize}
\tablecolumns{6}
\tablewidth{0 pt}
\tablehead{
\multicolumn{6}{c}{\textbf{CBC Injection Properties}} \\ [-0.1cm]
\colhead{Binary Type} & \colhead{Object} & $m/\msun$ (min/max) & $m$ distribution & $\chi$ distribution & max redshift}
\startdata
\ac{SSM} & $m_1$ & 0.2 - 15     & $m^{-2.5}$    & 75\% Uniform(0 - 0.4) + 25\% Uniform(0-1.0) & 0.25  \\ [-0.1cm]
         &  $m_2$ & 0.2 - 1.0   & $m^{-1}$      & 75\% Uniform(0 - 0.4) + 25\% Uniform(0-1.0) &       \\[0.1cm]
\ac{BNS} & $m_1$ & 1.0 - 3.0    & Uniform       & 75\% Uniform(0 - 0.4) + 25\% Uniform(0-1.0) & 0.25  \\ [-0.1cm]
         &  $m_2$ & 1.0 - 3.0   & Uniform       & 75\% Uniform(0 - 0.4) + 25\% Uniform(0-1.0) &       \\[0.1cm]
\ac{NSBH} & $m_1$ & 3.0 - 60.0  & $m^{-1}$     & Uniform(0 - 1)                              & 0.75  \\ [-0.1cm]
         & $m_2$ & 1.0 - 3.0    &  $m^1$        & 75\% Uniform(0 - 0.4) + 25\% Uniform(0-1.0) &       \\[0.1cm]
\ac{BBH} & $m_1$ & 3.0 - 1000   & $m^{-2.35}/ m^{-3}/ m^{-1}$   & Uniform(0 - 1)  & 2.5   \\ [-0.1cm]
         & $m_2$ & 3.0 - 1000   & $m^1$         & Uniform(0 - 1)  &       \\
\enddata
\caption{Distribution of intrinsic properties (component masses $m$ and spins $\chi$) of binary systems in the \ac{CBC} injection sample in LLPIC. The primary mass ($m_1$) for the \ac{BBH} population uses a broken power law distribution with breaks at 100 $M_\odot$ and 200 $M_\odot$. The spin distributions are isotropic in orientation.
\label{table:inj_dist}}
\end{deluxetable*}

\subsection{Burst Injections}

The Burst injections in the LLPIC data served a different purpose from the regular injection studies conducted by the \ac{LVK} search groups to characterize the sensitivity of Burst search pipelines~\cite{LIGOScientific:2025abo}.
In LLPIC, the number of Burst injections was limited to only 82. The injection times were chosen to test different detector network configurations (2 or 3 detectors observing simultaneously), to be spaced by more than one hour from each other and not to overlap a \ac{CBC} injection. This ensured that no single injection fell during interferometer downtime. Several waveform models were considered. First, generic waveforms were selected to describe the unmodeled nature of Burst sources. They include sine functions in a Gaussian envelope, band- and time-limited stochastic white noise bursts, and Gaussian pulses. Intrinsic parameters for these waveforms such as duration, frequency or bandwidth were varied to scan the search parameter space. In addition, astrophysically-motivated signals were injected: they include core-collapse supernova waveforms derived from numerical simulations~\cite{Mezzacappa:2022xmf,Pan:2020idl,Powell:2018isq} and phenomenological gravitational waveforms from cosmic strings~\cite{Damour:2000wa,Damour:2001bk,Damour:2004kw}. The injection amplitudes were randomly drawn in a range tuned to offer a good balance between significant detections and signals invisible to searches. The incoming sky directions of the injected \ac{GW}s were randomly distributed across a sphere of the sky.

\subsection{Results}

\subsubsection{Pipeline Performance}

The replay of LLPIC started from April 29, 2025.  A majority of the low-latency search pipelines (GstLAL, PyCBC, MBTA, cWB, MLy) participated in analyzing the LLPIC data during its fourth cycle, which ran from May 20 to 27, 2025. Previous cycles were used for testing data streaming and preparation for analysis. SPIIR and Aframe performed their analysis on different \ac{LLPIC} cycles. In the following, we study the \ac{GW} triggers
and their corresponding data products uploaded to \ac{GraceDB} during the $4^{th}$ cycle. Triggers and injections are cross matched using the merger time; all triggers within 1\,s of an injection merger time are matched to that injection and included in the analysis. This process removes noise triggers between injections and ensures trigger times correspond to an injection time window. This process, however, does not completely filter noise triggers and they can still be coincident with the injection time window. This cross matching process allows us to evaluate data products by comparing the results with the injected quantities in a similar manner to the O3 MDC \citep{O3_mdc}. 

During the $4^{th}$ cycle of LLPIC, 332 significant \ac{CBC} signals were cross-matched with the injections among which 45 were \ac{SSM} injections, 86 were \ac{BNS}, 63 were \ac{NSBH}, and 138 were \ac{BBH} injections. Despite an \ac{SNR} cutoff of 6, $\sim 75\%$ of \ac{CBC} signals are not recovered. Most of these signals were lost since they were injected during a time when one detector was not operational, or both were not operational or not in science mode. Other factors like large distance and possible glitch overlap also affect successful recovery of the signals.


\begin{figure}[h]
\centering
    \includegraphics[width=\columnwidth]{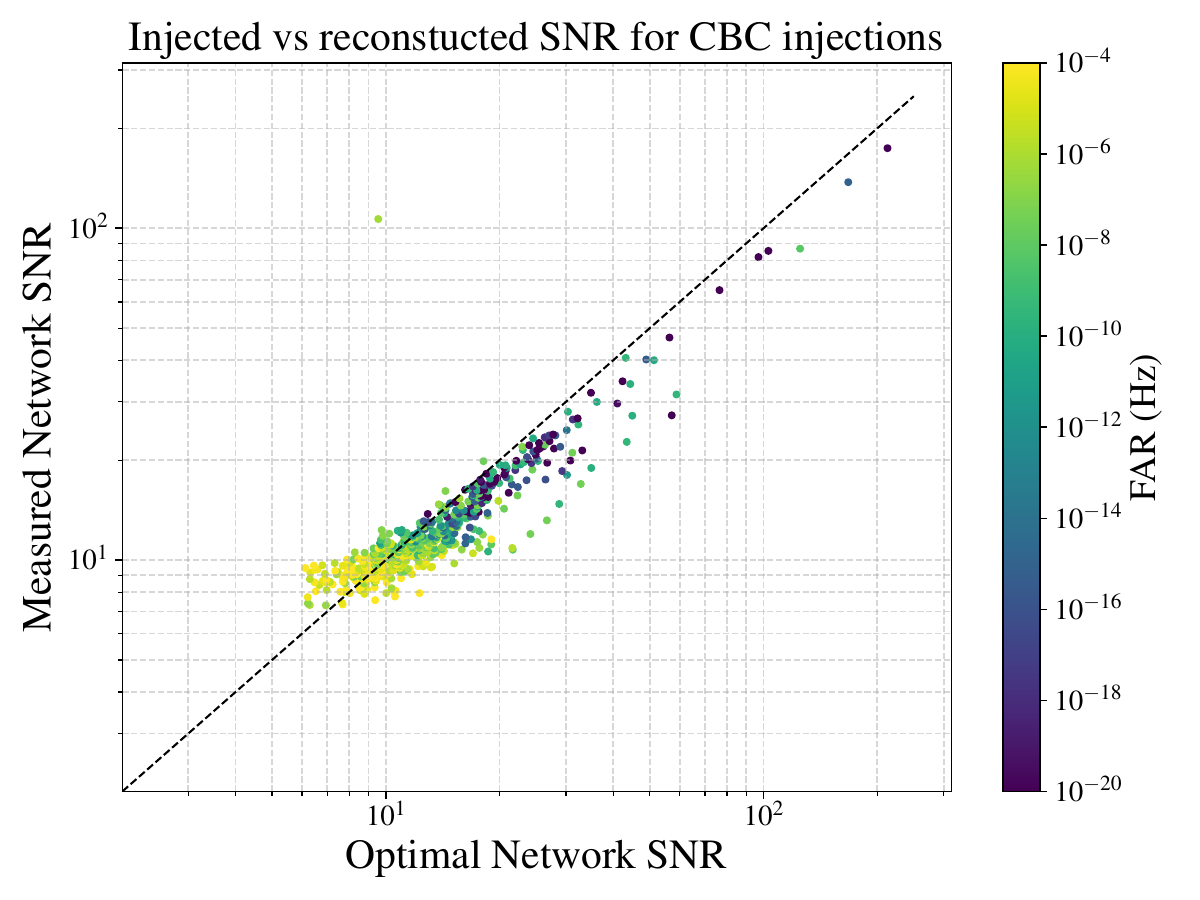}
    \caption{The measured network SNR recovered during the LLPIC cycle 4 compared to the optimal, or injected, network SNR with the points colored by FAR. }
    \label{fig:inj_rec_snr}
\end{figure}

The simulated and recovered network \acp{SNR} are compared in
Figure~\ref{fig:inj_rec_snr}. Signals of moderate \ac{SNR} are recovered
accurately, while a systematic bias is present at low \ac{SNR}, arising
from the \ac{FAR} threshold imposed for candidate upload, which
preferentially selects injections whose recovered \ac{SNR} fluctuates
above the true value. The loudest signals are recovered with marginally
lower \ac{SNR} than injected. This behavior is expected, as the optimal
network \acp{SNR} were computed assuming fixed detector sensitivities and
a complete three-detector network, whereas the observed data exhibit
time-dependent sensitivity variations and a subset of injections was
recovered by only two detectors or a single detector. 



\begin{figure}[h]
\centering
    \includegraphics[width=\columnwidth]{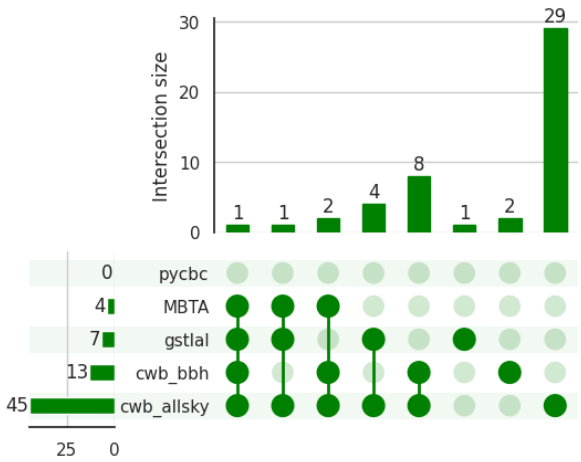}
    \caption{Overlap between Burst and CBC searches for Burst injections. The number of detected Burst injections is indicated for each search pipeline on the left side of the diagram. The histogram counts the number of detected injections for each combination of pipelines. The unmodeled Burst search using cWB is labeled ``cwb\_allsky'' while the \ac{BBH}-model version of cWB is labeled ``cwb\_bbh''.}
    \label{fig:llpic_burst}
\end{figure}

In case of Burst injections, 45 of them were detected as significant triggers by a Burst pipeline, \ac{cWB} or MLy. All of them were detected by cWB while MLy offered a redundant detection for 7 injections. Interestingly, Burst injections can also be detected by CBC pipelines. Fig.~\ref{fig:llpic_burst} presents the number of Burst injections detected by the different low-latency pipelines. The version of cWB configured to target \ac{BBH} signals, labeled ``cwb\_bbh'', detected 13 Burst injections. This result is expected as the pipeline is only mildly modeled for \ac{BBH}s and remains sensitive to non-\ac{CBC} signals. The \ac{CBC} matched-filter pipelines, MBTA and GstLAL, detected 4 and 7 Burst injections respectively showing that \ac{CBC} templates can trigger on Burst waveforms, including sine-Gaussian, cosmic string and supernova waveforms. These injections were all recovered with P(BBH)~$>96\%$. Indeed, Burst signals can be misinterpreted as heavy \ac{BBH}s with very few cycles in the bandwidth of the detectors~\cite{LIGOScientific:2020ufj}. The majority of these signals were detected with a higher significance by Burst pipelines. However one injection, a cosmic string signal, was best detected by GstLAL: the \ac{SNR} reported by GstLAL is 19.7 while the cWB \ac{SNR} is 18.2, and the false-alarm rate for this event is $3.0\times 10^{-9}$~Hz and $1.8\times 10^{-8}$~Hz for GstLAL and cWB respectively. This example shows that very short duration \ac{GW} candidates can be misidentified as \ac{CBC} and that careful follow-up studies must be performed. Fig.~\ref{fig:llpic_burst} also shows that the PyCBC pipeline did not detect any Burst injections, indicating that its templates are the least susceptible to triggering on non-\ac{CBC} transients among the participating \ac{CBC} pipelines.

\subsubsection{Latency Measures}

To enable prompt multi-messenger follow-up, a central goal of the \ac{LLAI} is the rapid dissemination of \ac{GW} alerts along with data products. In this study, we systematically examine the alert latency arising from three primary components of the pipeline: (i) the search pipelines, (ii) the event orchestrator \gwcelery, and (iii) \ac{GraceDB}. Latency contributions from data calibration, frame construction, and inter-site transfer -- typically taking about 5–10\,s -- are not included, nor are delays associated with alert ingestion and redistribution through \acs{GCN}\footnote{\url{https://gcn.nasa.gov/}} or \acs{SCiMMA}\footnote{\url{https://scimma.org/}}. The O3 MDC \citep{O3_mdc} has shown that the \ac{LLAI} system is capable of sending alerts at a median latency of $\sim$ 30\,s. We note that occasional technical issues during the \ac{LLPIC} can lead to unusually high-latency outliers; therefore, the latencies reported here represent conservative estimates, excluding the time required for strain data preparation and transfer.

\begin{deluxetable}{cCCC}
\tablehead{
\colhead{Latency Measure} & \colhead{Description} & \colhead{50\% (s)} & \colhead{90\% (s)}
}
\startdata
Superevents & \tsuperevent - \tmerger & 8.4 & 49.8 \\
CBC Events & \tevent - \tmerger & 8.9 & 62.9 \\
CBC Events (SNR Optimized) & \tevent - \tmerger & 129.4 & 231.6 \\
Burst Events (cWB-BBH) & \tevent - \tmerger & 70.8 & 123.6 \\
Burst Events (unmodeled) & \tevent - \tmerger & 90.6 & 166.0 \\
Early Warning Events & \tevent - \tmerger & -3.8 & 5.9 \\
GW Advocate Request & \tadvreq - \tmerger & 13.0 & 60.6 \\
GCN Preliminary Sent & \tgcn - \tmerger & 26.2 & 77.6 \\
\enddata
\caption{Measured latencies for a number of steps in the pipeline. $t_0$ corresponds to the event merger time reported by the pipeline, while $t_{superevent}$ and $t_{event}$ correspond to the time of superevent or event creation. For the case of superevent latencies, $t_0$ is determined by the preferred event.}
\label{table:latencies}
\end{deluxetable}

\begin{figure*}[ht!]
\includegraphics[width=0.5\linewidth]{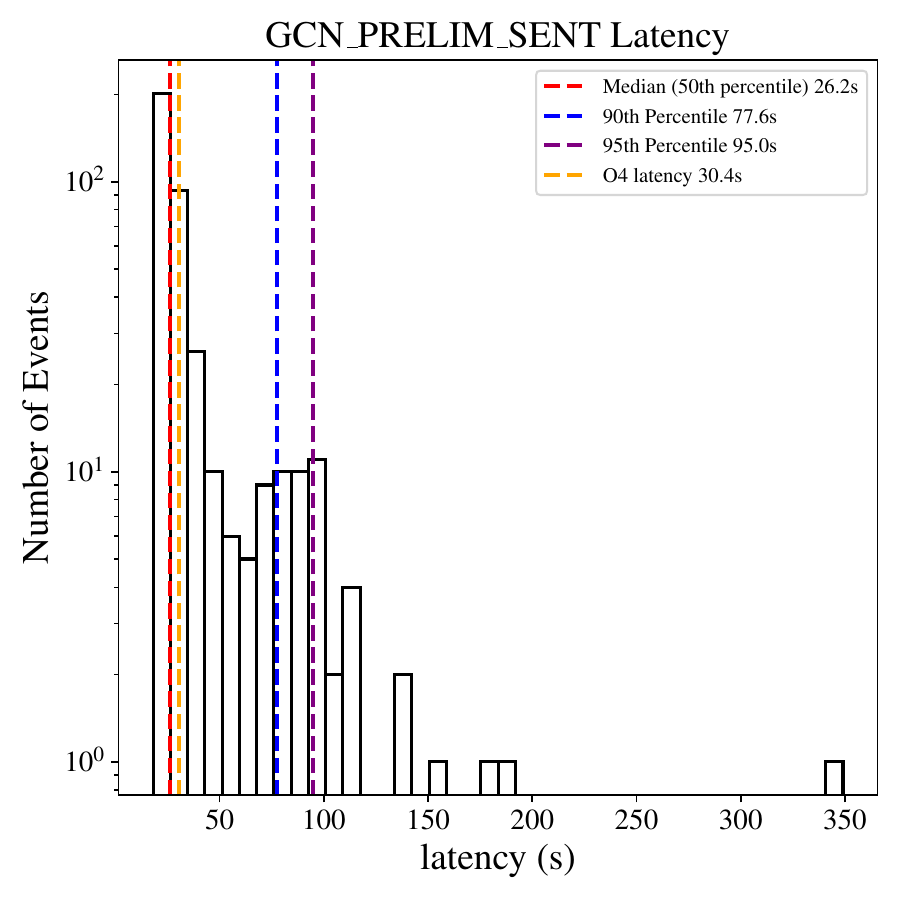}
\includegraphics[width=0.5\linewidth]{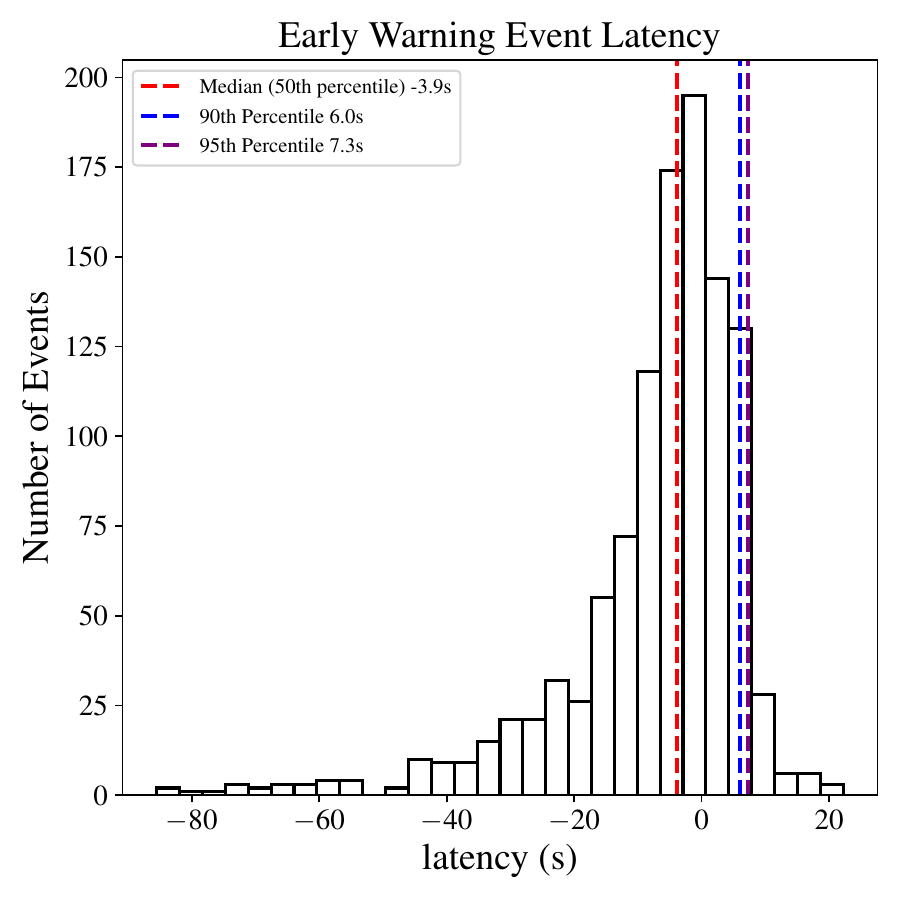}

\caption{\textit{Left}: Histogram of latencies for the sending of the GCN preliminary alert. We compare the \ac{LLPIC} latencies to the median latency measured during \ac{O4}. The \ac{O4} measurement includes data calibration, construction, and transfer time, while the \ac{LLPIC} latencies do not. \textit{Right}: Histogram of latencies for Early Warning events submitted to \ac{GraceDB}. Both low-significance and significant events were considered for the Early Warning plot.}
\label{fig:latencies}
\end{figure*}

Event latencies are computed by comparing the time an event is created and appears on \ac{GraceDB} to that of the known merger time. For \ac{GW} event candidates during an observing run, the merger time is defined as the time the signal peak reaches Earth's center.

For \ac{CBC} pipelines, we find a median (90\%) latency of 8.9\,s (62.9\,s); for Burst unmodeled pipelines, we find a median (90\%) latency of 90.8\,s (166.0\,s). During \ac{O4} GstLAL and \ac{MBTA} implemented an SNR-Optimized search \citep{Joshi:2026iux, Aubin_2026} which is a refinement over the first \ac{CBC} trigger. The median latency for the \ac{SNR}-Optimized triggers stands at 129.4 s which makes them a candidate for the second preliminary alert.  For Early Warning alerts, shown on the right of Figure~\ref{fig:latencies}, we find a median (90\%) latency of -3.9\,s (6.0\,s).

Within \gwcelery ~the median (90\%) time for creation of a superevent is 8.4\,s (49.8\,s). The median superevent latency is lower than the event latency due to the fact that the superevent may be created upon the first trigger, and that a superevent often consists of multiple events. 
Once a significant superevent has been created, there is a request for human vetting of the alert and the start of a rapid response procedure, called the Advocate Request; we find a median (90\%) latency of 13.1\,s (60.6\,s) to notify the advocate. To measure the latency of event communication to the community, we also measure the latency for sending of the GCN preliminary alert, which occurs for superevents that pass the significance threshold; we find a median (90\%) latency of 26.2\,s (77.6\,s) for full bandwidth events.  We show this statistic for GCN preliminary alerts on the left of Figure~\ref{fig:latencies}. This latency reported specifically measures the time until the GCN preliminary label is applied. During the entire one week cycle, 23 significant early warning events were found and we found the median latency for public early warning alerts to be 6.95 s. Some early warning events were sent with negative latency with the earliest being -30.8 s.   

We also compare this to the measured median GCN latency during \ac{O4} which, including data calibration, data aggregation, and transfer time, is 30.4\,s. The difference between GCN latency during the \ac{LLPIC} and \ac{O4} is due to the fact that real \ac{O4} data includes data calibration, construction, and transfer time latency, while the \ac{LLPIC} measurement does not. This is smaller than the previously reported \ac{MDC} number of 29.5 s \cite{O3_mdc} due to various changes made in \ac{GraceDB} and \texttt{igwn-alert}. Table \ref{table:latencies} shows a compilation of these latency statistics for comparison.

\subsubsection{Probability of astrophysical origin}
\label{subsec:pastro_results}

\begin{figure}[ht!]
\centering
\includegraphics[width=\columnwidth]{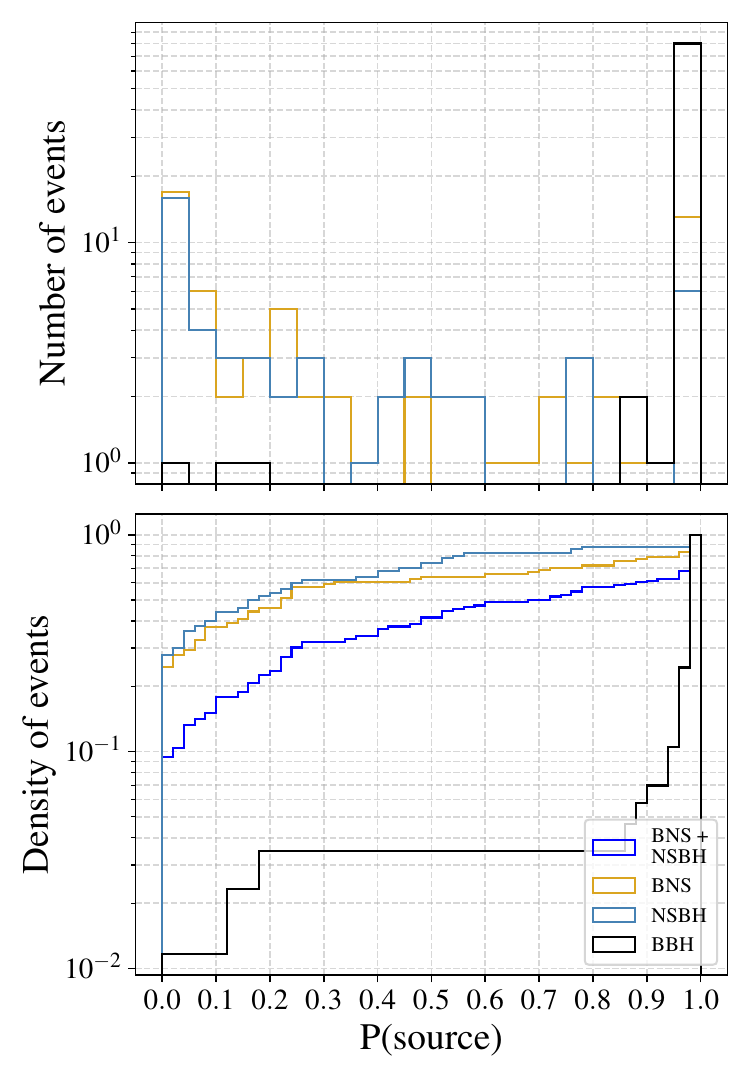}
\caption{\textit{Top:} Histogram of ``preferred'' recovered P(source) for true sources, for superevents which pass the significant public alert threshold.
The possible source classes are \ac{BNS}, \ac{NSBH}, and \ac{BBH}.
This excludes early warning events, which were not fully functional during the time of this analysis.
\textit{Bottom:} Cumulative density of the same data. We also include the distribution of correctly recovered \ac{BNS} or \ac{NSBH}, which checks for contamination between the two classes due to misrecovered secondary masses or effects from varying the \ac{NS}/\ac{BH} mass boundary. As this is a cumulative histogram, the fraction of events above a certain P(source) corresponds to the \ac{TPR}. 
\label{fig:psource}}
\end{figure}

\pastro is defined as the probability that a particular \ac{GW} trigger is
caused by an astrophysical signal as opposed to the probability of being caused by a noise fluctuation or terrestrial contamination, $p_{\rm terrestrial}$. \pastro is further divided into probabilities that the binary is a \ac{BNS},
\ac{NSBH} or \ac{BBH} given by $p_{\rm BNS}$, $p_{\rm NSBH}$, and $p_{\rm BBH}$, respectively, to provide more information regarding the \ac{GW} source class.
Each \ac{GW} superevent inherits the \pastro estimate from its preferred \ac{CBC} event. Burst unmodeled searches do not provide an estimate of \pastro. The \pastro mass parameter space has not yet been expanded onto \ac{SSM} regions and thus \ac{SSM} \ac{CBC} searches report an empty \pastro.

By matching \ac{LLPIC}'s injected parameters to the recovered \pastro values from the superevent, we can test the accuracy of \pastro.
Figure~\ref{fig:psource} shows the recovered P(source) for true sources (e.g., P(BNS) for true injected \ac{BNS} events), for superevents which pass the public alert FAR threshold.

In matching these injections, we place the cut between \ac{NS} and \ac{BH} at 3 $M_\odot$. Injected \ac{BBH}s are typically recovered confidently, with the vast majority resulting in P(BBH) $>$ 0.5. Over 90\% of \ac{BBH}s are recovered with P(BBH) $>$ 0.9.
The P(BNS) and P(NSBH) distributions are less confident than the P(BBH) distribution; about 20\% of true \ac{BNS}s and \ac{NSBH}s are recovered with P(BNS) or P(NSBH) $< 0.1$.
As a check against contamination across P(BNS) and P(NSBH) due to errors in recovered masses or variation in the definitions of the mass border between \ac{NS} and \ac{BH} (i.e., the Tolman–Oppenheimer–Volkoff mass), we can also look at the distribution of the sum of P(BNS) and P(NSBH) for injected events where $m_2 \leq 3 M_\odot$. This is shown in the bottom panel of Figure~\ref{fig:psource}. 

Compared to P(BNS) or P(NSBH) alone, P(BNS) + P(NSBH) performs better, with $\sim 40\%$ of true \ac{BNS} or \ac{NSBH}s receiving P(BNS) + P(NSBH) $> 0.9$, compared to 10\% and 20\% for \ac{NSBH} and \ac{BNS} respectively. If instead we use a threshold P(source) of $0.5$, we find a \ac{TPR} of $\sim 98$ \% for \ac{BBH}, $\sim 35$ \% for \ac{BNS}, and $\sim 25$ \% for \ac{NSBH} injections. 


\subsubsection{Sky Localization}

One of the key data products to enable multi-messenger follow-up is the rapid inference of the sky localization from \ac{GW} observations. This sky localization consists of the posterior probability distribution of the source location in the sky. The sky localization, mapped either over a 2D map of right ascension and declination, or a 3D volume which also includes a distance estimate, is known as a ``sky map.'' Sky localization (and parameter estimation more generally) is conducted in multiple stages once a candidate is identified. For \ac{CBC} sources, \ac{BAYESTAR}, a rapid sky localization algorithm \citep{SiPr2016}, is used to generate sky maps, and may be updated by \texttt{Bilby}, a Python-based parameter estimation pipeline that uses stochastic sampling methods \citep{2019ApJS..241...27A, Romero-Shaw:2020owr}. Sky maps from \acs{BAYESTAR} are released with Preliminary Notices and sky maps from \texttt{Bilby} are released in Update Notices. Burst pipelines, on the other hand, produce their own sky maps and upload them to \ac{GraceDB}.

In order to evaluate localization performance of the sky maps, the localization area and searched area are computed.   Localization area is the area (measured in$~\mathrm{deg^2}$) which encloses a given probability contour in the sky map; in this work, we use 90\% as the total cumulative probability threshold. Searched area is the smallest 2D area, starting with the regions of highest probability, that contains the true location of the source; it represents a measurement of the sky area that a telescope with a small field of view relative to the sky map size would need to cover before imaging the true location. We refer the reader to \cite{SiPr2016} for more details on these parameters, and use the \texttt{ligo.skymap}\footnote{\url{https://git.ligo.org/leo-singer/ligo.skymap}} package to compute all metrics. 

When evaluating sky map performance, we consider the preferred events for superevents that fall under the significant FAR threshold before trials factor and were detected by more than one interferometer. Most search pipelines require a candidate to be identified above threshold in at least two detectors, but some pipelines like \ac{GstLAL} can also report candidates found above threshold in only one. For these single-detector candidates, \ac{BAYESTAR} can still use the SNR time series from any other observing detectors, even though the signal was below threshold there. If only one detector was observing, the localization is set by that detector’s antenna pattern alone and covers a large fraction of the sky regardless of SNR. In this study, we exclude such single-detector triggers, since \ac{BAYESTAR} was not run with data from multiple detectors and their localizations alone are not meaningful for assessing localization performance.

\begin{figure*}[ht!]
\centering
\includegraphics[width=0.49\linewidth]{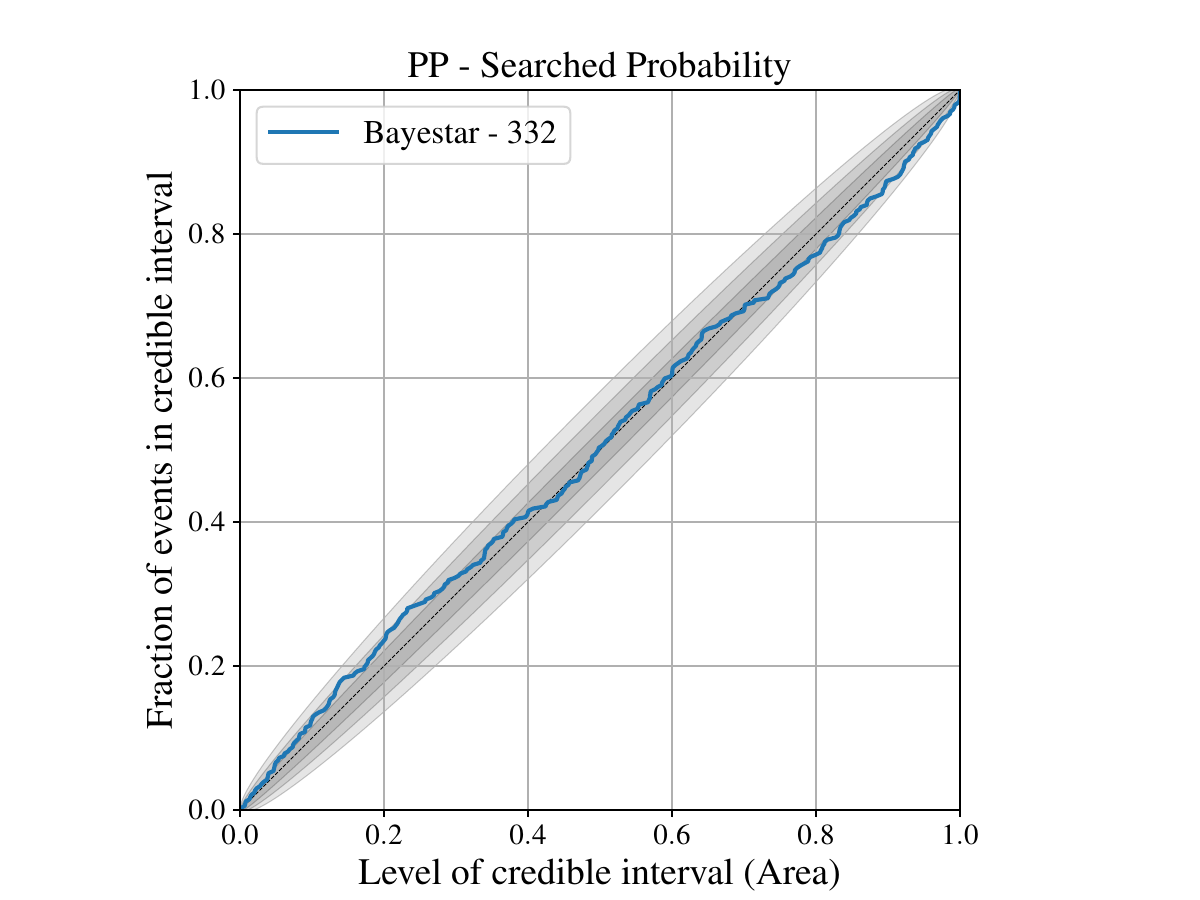}
\centering
\includegraphics[width=0.49\linewidth]{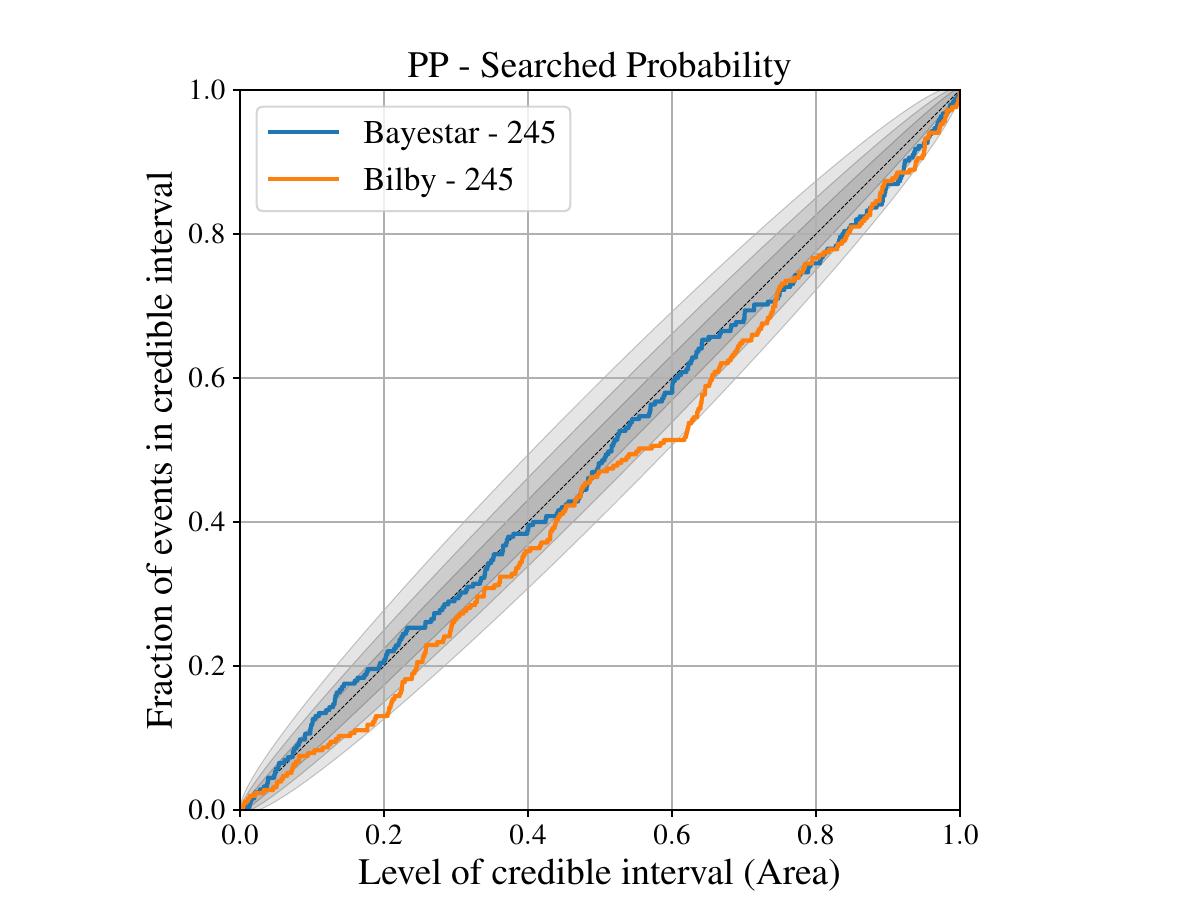}
\centering
\includegraphics[width=0.49\linewidth]{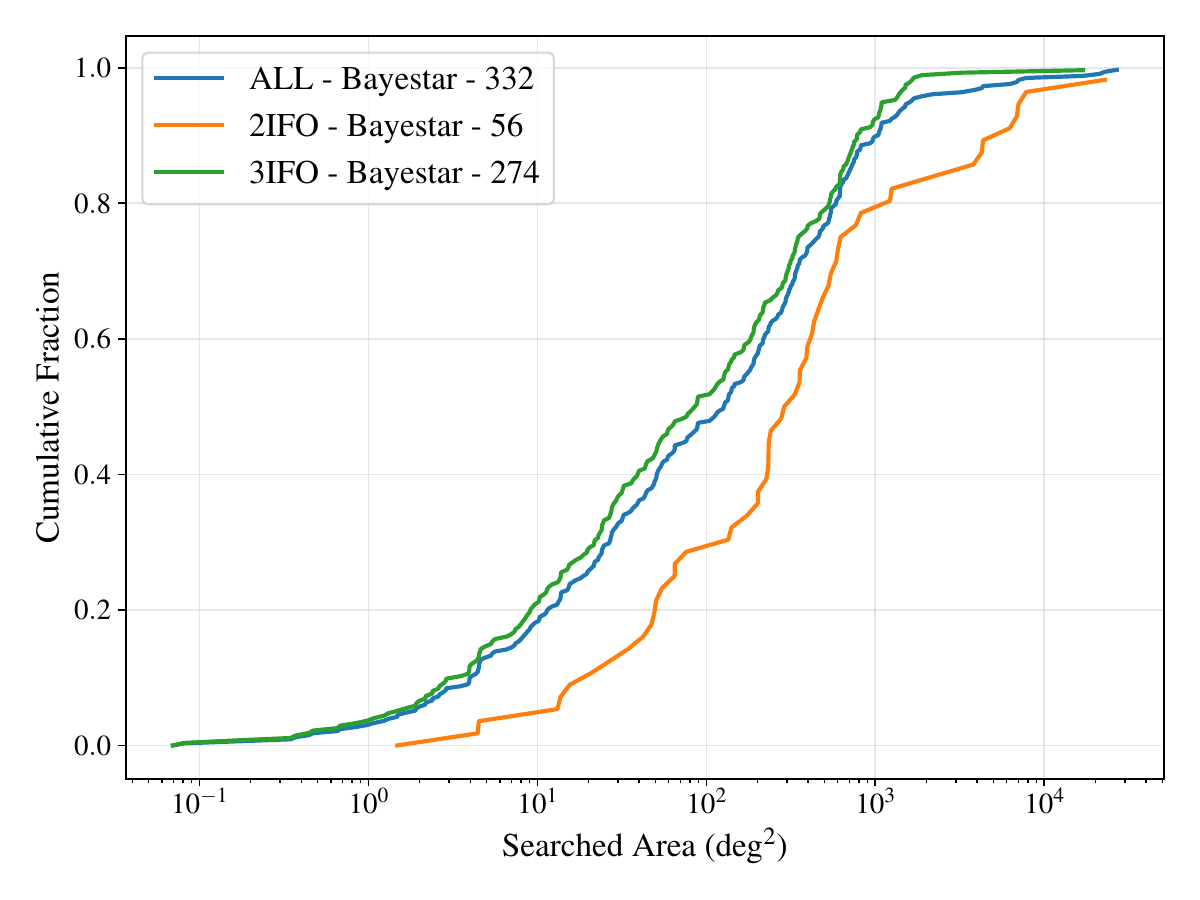}
\centering
\includegraphics[width=0.49\linewidth]{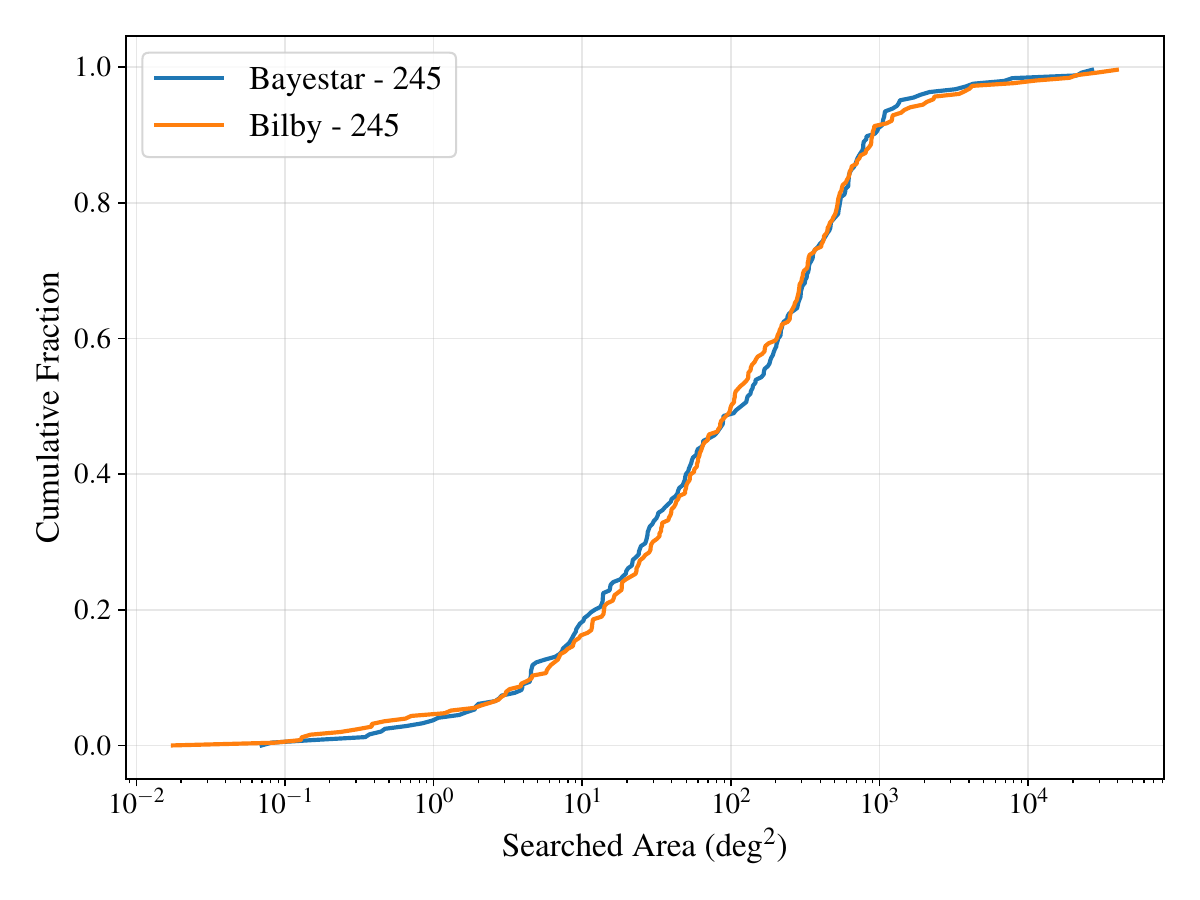}
\caption{\textit{Top left:} A \textit{P-P} plot showing the \texttt{BAYESTAR} sky map statistics for the preferred event. The credible intervals shown in gray are based on the total number of events. \textit{Bottom left:} Cumulative histograms of \texttt{BAYESTAR} searched area for all events (blue), compared to two-interferometer (Orange), three-interferometer (Green). 
\textit{Top right:} A \textit{P-P} plot showing the performance of \texttt{BAYESTAR} (blue) and \texttt{Bilby} (orange) generated sky maps for \ac{BNS} events. Preferred events where both \texttt{BAYESTAR} and \texttt{Bilby} sky maps are available were considered for the plot. \textit{Bottom right:} Cumulative histograms showing searched area statistics for \texttt{BAYESTAR}  and \texttt{Bilby} sky maps.}
\label{fig:skymaps}
\end{figure*}


Further, in the top left panel of Fig.~\ref{fig:skymaps}, we show the accuracy of \texttt{BAYESTAR} sky maps through a \textit{P-P} plot. Sky maps corresponding to the preferred event of a superevent that passes the current CBC FAR threshold of 1/6 month were considered. \textit{P-P} plots of this format show the fraction of injections found within a given credible interval across all levels of credible intervals. The three gray regions around the diagonal show three different levels of confidence (1-3 $\sigma$) for the combined \texttt{BAYESTAR} map sample. We find that the \texttt{BAYESTAR} sky maps fall well within the credible intervals which implies a good accuracy.  In the bottom left panel of Fig.~\ref{fig:skymaps}, we show the cumulative trend of the searched area statistics from the combined sample. We find a median searched area of $~100-200~\mathrm{deg^2}$. We then compare the two interferometer events with the three interferometer events, to show that the latter produced smaller searched areas. The numbers in the legend indicate the total number of events used in the plot.  

Further, we probe the accuracy of the \ac{BAYESTAR} sky maps compared with \texttt{Bilby} sky maps. Due to the dense rate of injections and limited resources, not all PE runs were completed for significant superevents in \ac{LLPIC}. The total number of Bilby runs crossmatched with injections during the fourth \ac{LLPIC} cycle was found to be 245. We compare the \textit{P-P} plot and searched area histogram on right side of Fig.~\ref{fig:skymaps} for the preferred event of all superevents for which both sky maps were produced. From the top right panel of Figure~\ref{fig:skymaps}, we see that both the \texttt{BAYESTAR} and \texttt{Bilby} sky maps recover the injected source locations within their credible intervals at close to the expected rate across all confidence levels.  The cumulative searched area plot in the bottom right panel of Fig.~\ref{fig:skymaps} shows that both \ac{BAYESTAR} and \texttt{Bilby} sky maps have comparable searched area.  In the O3 Replay \ac{MDC}, we saw that Bilby sky maps tend to have lower searched area compared to \ac{BAYESTAR} \citep{O3_mdc}. The comparable cumulative searched area is likely due to higher-\ac{SNR} events than in the O3 \ac{MDC}. The median \ac{SNR} of events for which Bilby runs were completed is $\sim$14.

\subsubsection{EM-Bright}
EM-Bright is a pipeline designed to assess binary source properties and whether a \ac{GW} candidate is capable of producing an electromagnetic counterpart \citep{ChGh2020}. The \hasns, \hasremnant, \hasgap, and \hasssm~ data products are defined in Sec.~\ref{SSM_section}. EM-Bright quantities are reported as part of the automated and update discovery notices.

To assess the performance of the EM-Bright data products in \ac{LLPIC}, we compute their \ac{ROC} curves. The \ac{NS}--\ac{BH} boundary is set according to the SLy \ac{EoS}, but the probabilities are \ac{EoS}-marginalized, meaning that some uncertainty in the \ac{NS} \ac{EoS} is included in the classifications. The markers indicate three different representative thresholds (a score above which events are considered to be positively classified as the source type in question) along the \ac{ROC} curves for each pipeline. 

\begin{figure*}
\centering
\includegraphics[width=0.49\linewidth]{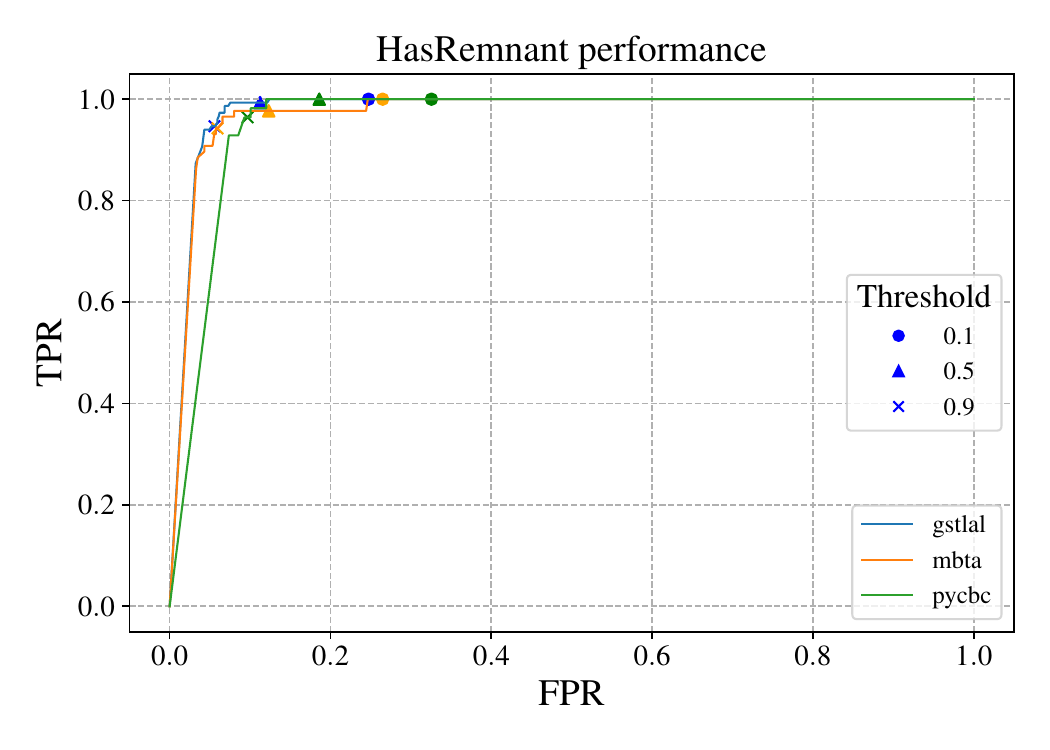}
\centering
\includegraphics[width=0.49\linewidth]{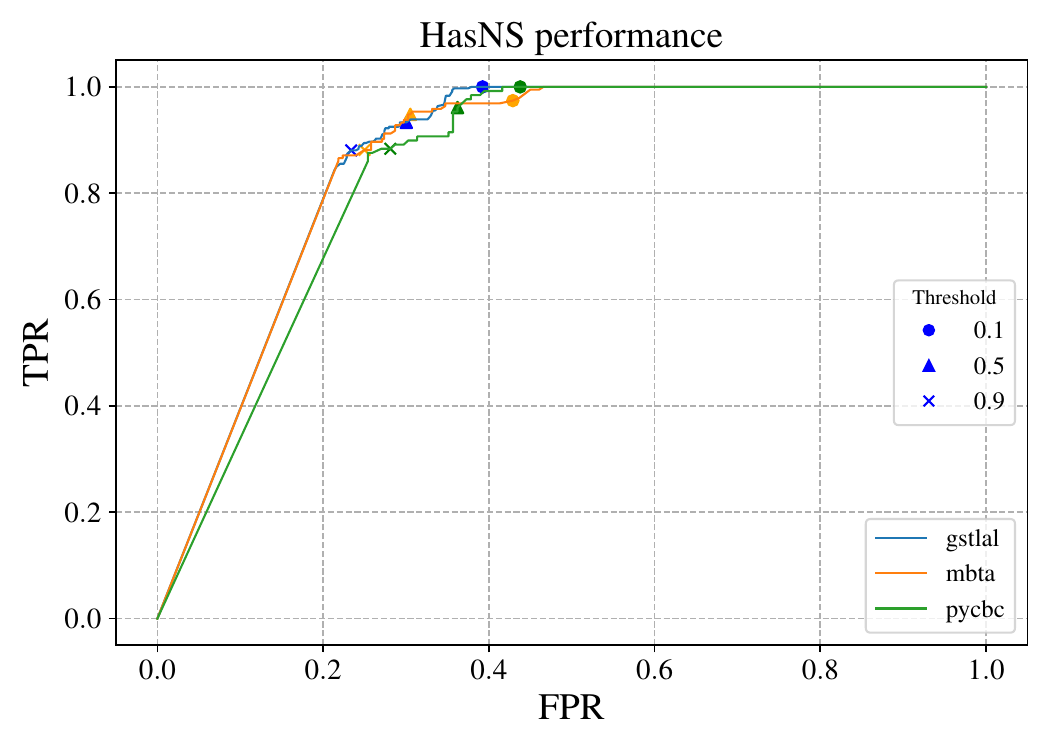}
\centering
\includegraphics[width=0.49\linewidth]{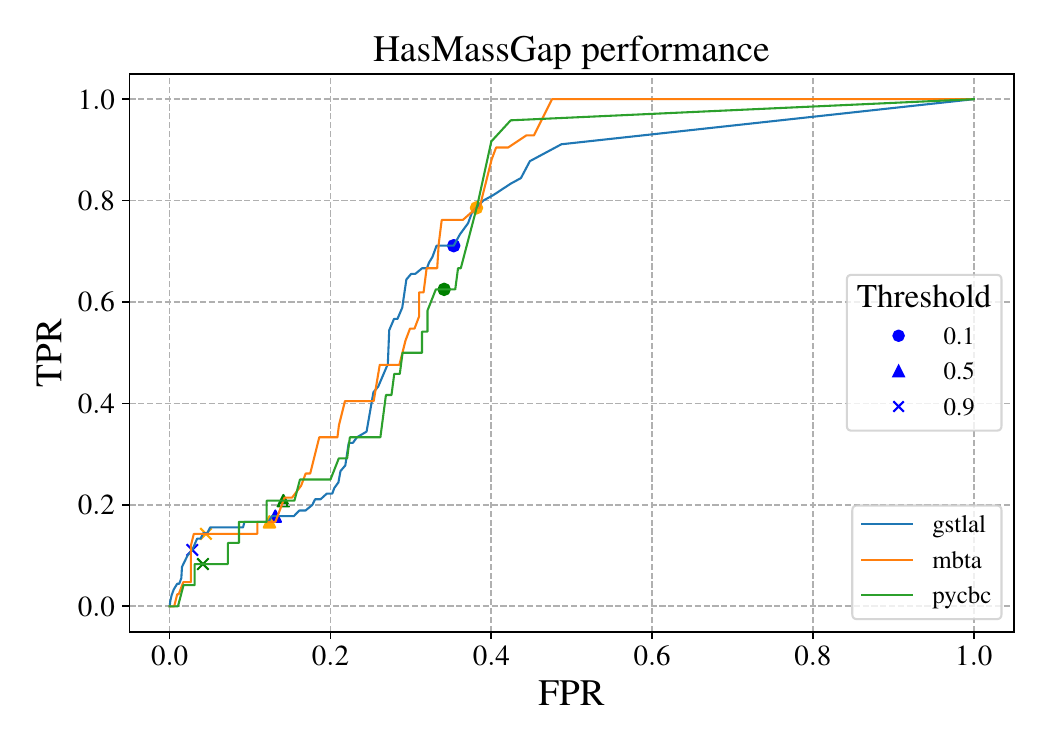}
\centering
\includegraphics[width=0.49\linewidth]{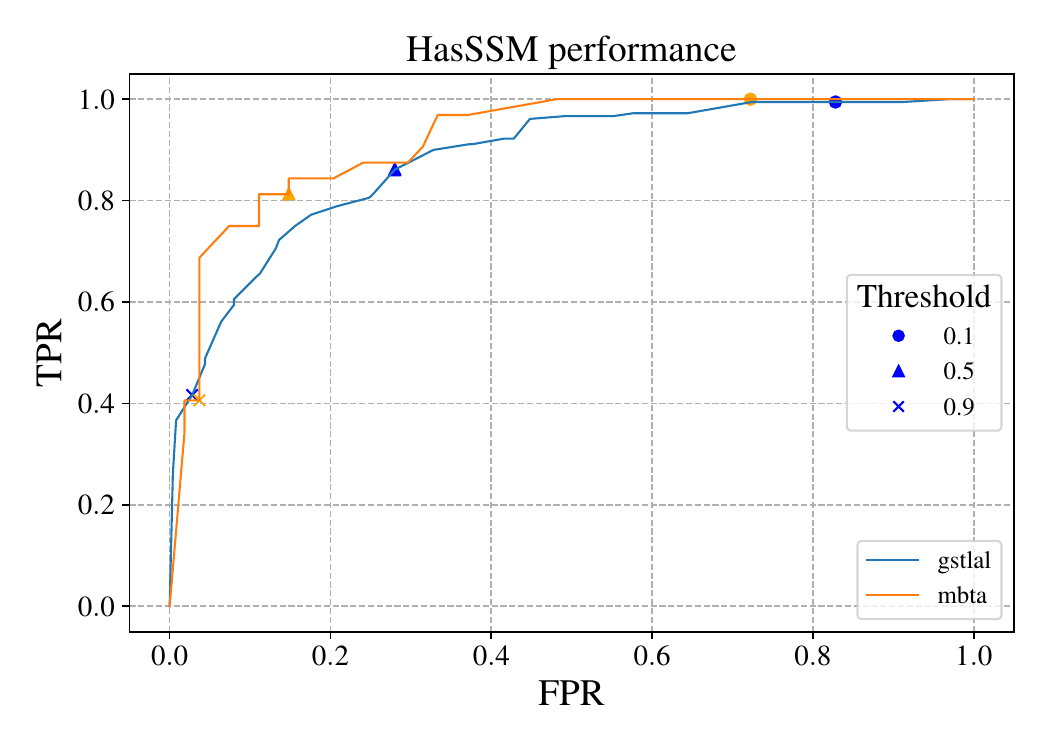}
\caption{The ROC curves for the different EM-Bright
classifiers are shown here for LLPIC events. The  {\hasremnant}, {\hasns}, and {\hasgap} performances are evaluated on regular \ac{CBC} alerts whereas {\hasssm} is computed for \ac{CBC} \ac{SSM} searches only. The markers denote different representative thresholds along the curve. }
\label{fig:sourceproperties}
\end{figure*}

In Fig.~\ref{fig:sourceproperties},  we see that the {\hasremnant} quantity for all participating \ac{CBC} pipelines has greater than a $\sim$ 90\% \ac{TPR} for a $\sim$ 10\% \ac{FPR}. In case of \hasns, we see that it performs at $\sim 90\%$ \ac{TPR} but at  $\sim 30\%$ \ac{FPR}. The \ac{FPR} is significantly higher than what was observed in the O3 Replay \ac{MDC}. Similarly, for {\texttt{HasMassGap}},  we see a drop in performance with $\sim 80\%$ \ac{TPR} at $\sim 40\%$ \ac{FPR}. 

\begin{figure}
\centering
    \includegraphics[width=\columnwidth]{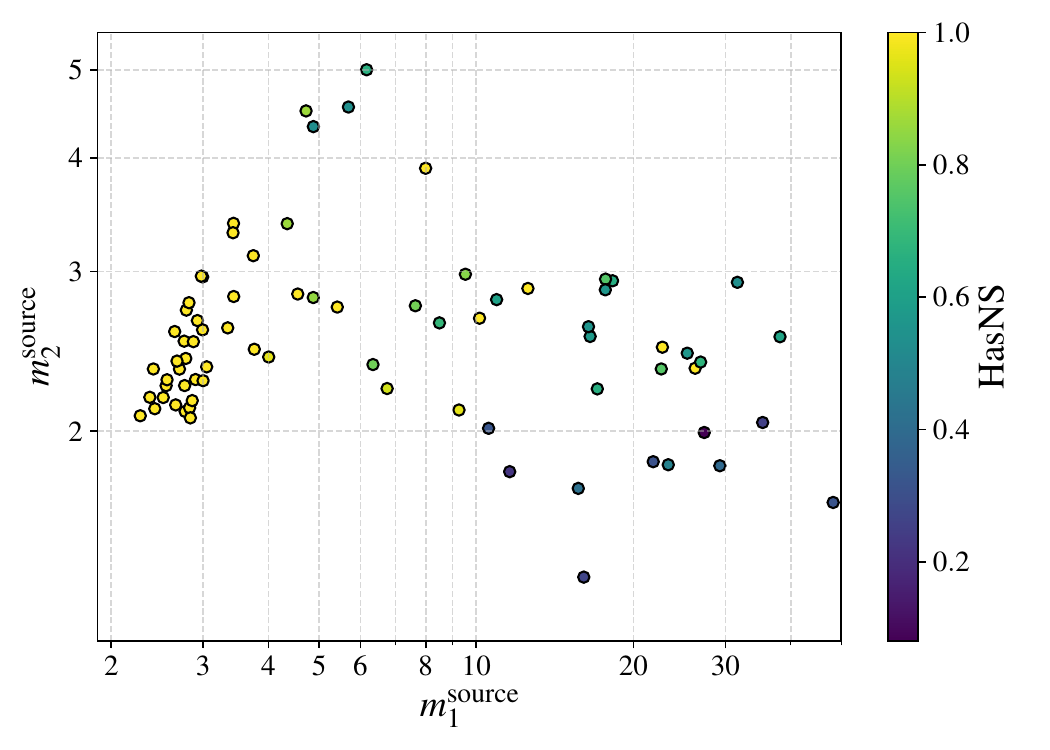}
    \caption{Primary and secondary masses of significant LLPIC triggers with misclassified HasNS.}
    \label{fig:hasns_explain}
\end{figure}

Fig.~\ref{fig:hasns_explain} is a scatter plot of source $m_1$ and $m_2$ depicting misclassified triggers. A trigger is considered misclassified if its \hasns\ score incorrectly crosses the decision threshold of 0.5 -- that is, when the score is below 0.5 for a true \ac{NS} injection (ground truth = 1) or above 0.5 for a non-\ac{NS} injection (ground truth = 0). We can see that the majority of the triggers that are misclassified are between 2-3 $M_\odot$. Since the SLy \ac{EoS} is considered as truth for classifying a \ac{NS}, all the injections above 2.05 $M_\odot$ are not considered as \ac{NS}. This causes an increase in \ac{FPR} which is clearly shown by the \ac{ROC} curve. If we choose a stiffer \ac{EoS} with maximum \ac{NS} masses around 2.8, the \ac{FPR} reduces by $\sim$ 20\% indicating that the classifier's performance near the \ac{NS} boundary depends on selection of \ac{NS} EoS. In the O3 replay MDC, the \ac{NS} injection boundary was set to 2.05 $M_\odot$ which resulted in fewer triggers in the region between 2-5 $M_\odot$ and the performance was found to be better compared to \ac{LLPIC}. However, since the classifiers were trained on an O2 era dataset, this drop in performance suggests retraining the classifier with the latest dataset might help in improving the classification performance. 

\subsubsection{Chirp mass}
\begin{figure}[h]
\centering
    \includegraphics[width=\columnwidth]{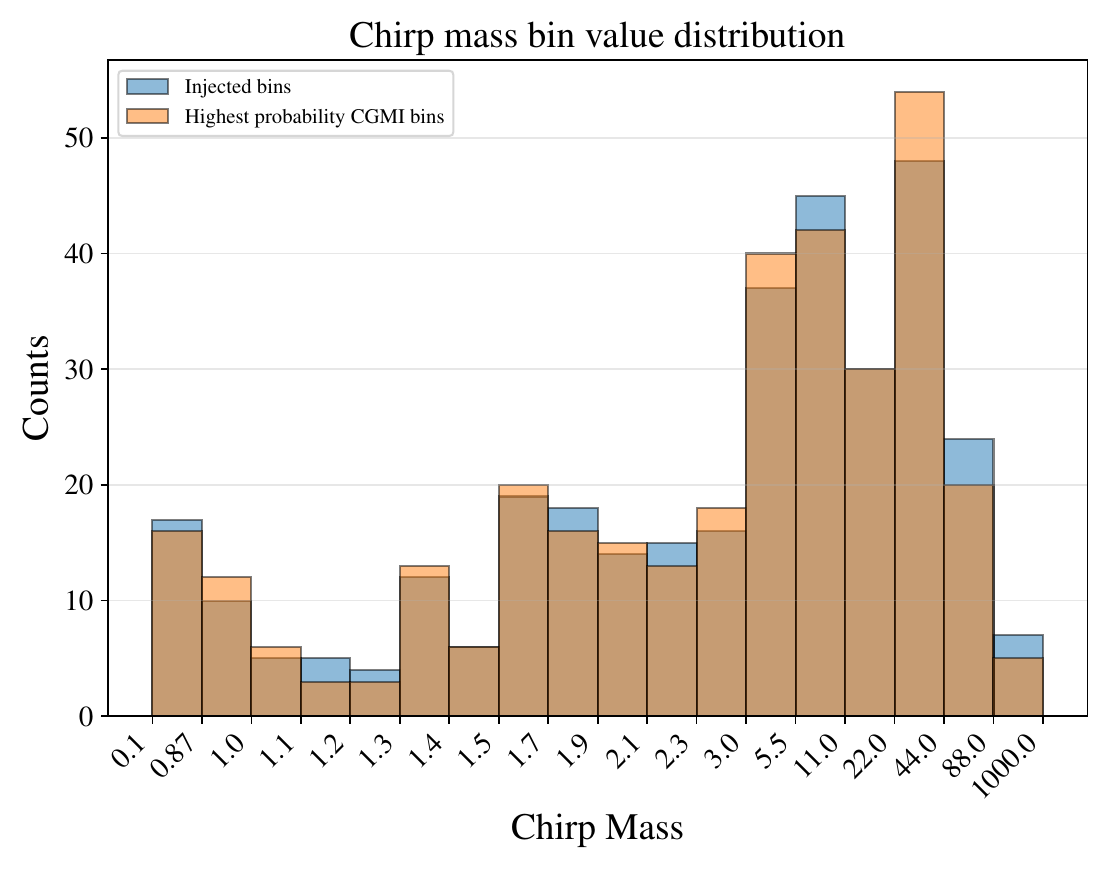}
    \caption{ Comparison of the most probable chirp mass bin distribution from low-latency estimates with the injected chirp mass distribution.}
    \label{fig:cgmi_llpic}
\end{figure}

Fig. \ref{fig:cgmi_llpic} shows a comparison of the highest probability chirp mass bin distribution based on low-latency estimates with the injected chirp mass for the significant \ac{CBC} events found in \ac{LLPIC} cycle 4. Ideally, the highest probability chirp mass bin distribution should match the injection chirp mass distribution.  We see that the mismatch between the two is limited to a small number of events in each bin. This shows that the low-latency binned chirp mass estimates are performing fairly accurately for the \ac{LLPIC} dataset.

\section{Conclusion}
\label{sec: conclusion}

This work summarizes key advances made in the \ac{LLAI} during \ac{O4} of the \ac{LVK}. We introduced the deployment of two new low-latency public data products -- notably \hasssm, a machine learning classifier quantifying the probability that the binary contains an SSM compact object, and binned chirp mass estimates. Both enhancements significantly expand the astrophysical and multi-messenger scope of \ac{LVK}'s real-time alert infrastructure.

We also describe the \ac{LLPIC} -- a comprehensive, end-to-end validation framework designed to test and strengthen the \ac{LVK}'s \ac{LLAI} ahead of the next observing runs. \ac{LLPIC} integrates realistic simulated \ac{CBC} and Burst \ac{GW} signals into archived O4 data and replays them through the entire alert generation pipeline. This approach enables systematic testing of new data products, latency performance, and cross-pipeline consistency under production-like conditions.

Public, low-latency chirp mass estimates are a significant step forward for enabling multi-messenger searches, as the mass information they provide can be used to characterize the source and possible counterparts. We intend for these mass estimates, which are provided in the form of binned probabilities and plotted and uploaded to \ac{GraceDB}, to be read alongside source classification by all consumers of \ac{LVK} alerts to gain a clearer picture of the candidate events we detect and enable both \ac{ToO} searches and downstream scientific efforts. S250818k has already demonstrated the potential usefulness of this data product by highlighting the standout nature of this candidate event, and more broadly, by estimating that it is likely a low-mass \hasns-type event, if it is in fact astrophysical. In the future, we expect these mass estimates to identify many more candidates of interest, whether they be low mass \ac{BNS} or high mass \ac{BBH}.

Performance analyses on \ac{LLPIC} show that low-latency data products  (\hasns~ and \pastro) show reduced performance compared to the previous \ac{MDC}. However, this can 
be attributed to the peculiar injection distribution of \ac{LLPIC}. Tuning these data products more in the \ac{NS}-\ac{BH} boundaries would help increase the performance in the upcoming observing runs. Other data products like sky-localization (\texttt{BAYESTAR} and \texttt{Bilby}) and the new chirp mass estimates are performing as expected for O4 sensitivity. Compared to O3 Replay MDC, we see a slight decrease in the median latency for preliminary alerts attributed to continuous developments in the \ac{GraceDB} and \texttt{igwn-alert} frameworks. Burst–CBC overlap tests further underscore the need for careful classification of short-duration transients, which can be misidentified across search pipelines.

We also developed and integrated the \ac{MEG}, a software package that replays previously recorded GW triggers and their associated data products through \ac{GraceDB} and \gwcelery ~(see appendix \ref{meg}). Within the \ac{LLPIC} framework, \ac{MEG} creates replay events that downstream infrastructure processes and the validation workflow assess, without rerunning the search pipelines. Its integration provides a foundation for continuous integration–style testing of new alert features before deployment. These developments mark a crucial step toward a more automated, reliable, and scientifically rich \ac{GW} alert infrastructure as the \ac{LVK} prepares for O5 and future observing runs.

\begin{acknowledgments}

We thank all the \ac{CBC} and Burst pipeline team for their contributions and for allowing the use of their pipeline data. A.T. and M.W.C. acknowledge support from the NSF with grant numbers PHY-2117997, PHY-2308862 and PHY-2409481. AT acknowledges support from the Director's Postdoc Fellowship at Los Alamos National Laboratory. S.S.C acknowledge support from the National Science Foundation with grant numbers PHY-2011334 and PHY-2308693. M.S. acknowledges Polish National Science Centre Grants No. UMO-2023/49/B/ST9/02777 and No. UMO-2024/03/1/ST9/00005, and the Polish National Agency for Academic Exchange within Polish Returns Programme Grant No. BPN/PPO/2023/1/00019. M.D. acknowledges the support of the Sapienza Grants No. RM123188F3F2172A and RG1241905E47F60C. D.S. acknowledges support from NSF Grant PHY-2020275(Network for Neutrinos, Nuclear Astrophysics, and Symmetries (N3AS)). S.K. acknowledges the NSF grant PHY-2409372 to the University of Florida. T.D. acknowledges support from grant PID2024-160643NB-I00 of the Ministerio de Ciencia, Innovaci{\'o}n y Universidades and Agencia Estatal de Investigaci{\' o}n, from Mar{\' i}a de Maeztu grant CEX2023-001318-M funded by MICIU/AEI/10.13039/501100011033, and from Grant ED431F 2025/04 of the Galician Conselleria de Educacion, Ciencia, Universidades e Formacion Profesional. This material is based upon work supported by NSF's LIGO Laboratory which is a major facility fully funded by the National Science Foundation. The authors are grateful to the LIGO Scientific Collaboration (LSC) for access to the O4b data on which this study is based, and for the computing resources used. 
\end{acknowledgments}

\appendix

\section{Mock Event Generator} \label{meg}

Over long observing runs, continuous development and integration of new features, search pipelines, and
data products are essential. However, the complexity of the system makes end-to-end testing challenging.
The O3 Replay MDC helped address this by providing a cyclic 40-day data stream processed by production-equivalent pipeline instances. Triggers were uploaded to a dedicated internal GraceDB instance named “playground”, with a parallel GWCelery deployment managing event processing and annotations. This configuration enabled realistic end-to-end validation of the low-latency system. However, maintaining such parallel pipeline instances over extended runs is computationally expensive. The subsequent development of the \ac{MEG} software package, described below, provided a more scalable and efficient approach to repeated testing of downstream \ac{LLAI} components.
\ac{MEG} is a software package developed by the \ac{LVK} to replay previously recorded generated \ac{GW} triggers in real time. It reproduces event submissions to \ac{GraceDB}, including sky maps and associated JSON data products, enabling downstream event processing and alert generation to be tested without rerunning the search pipelines

\ac{MEG} ensures that these signals appear as \textit{live alerts} to downstream infrastructure. This design allows for systematic validation of every stage of the workflow -- from the search pipelines, through \ac{GraceDB} ingestion, to the orchestration layer in \gwcelery. A key advantage is that \ac{MEG} provides reproducibility: identical events can be replayed across multiple cycles, allowing direct comparison of data product consistency. 

As part of its expanded functionality, \ac{MEG} introduced a structured review process to ensure the reliability and scientific integrity of replayed events. Each replay undergoes a series of automated and manual validation steps designed to mirror production-level quality control. These include verifying event metadata, such as ensuring that JSON fields like $p_{astro}$, \hasns, and binned chirp mass probabilities (see Sec.~\ref{CGMI}) are correctly populated, along with cross-checking sky maps generated by \ac{BAYESTAR} against injected parameters to confirm statistical consistency. The framework also performs timing and latency assessments to ensure that alert generation meets the \ac{O4} design goal of publishing initial alerts within tens of seconds of merger. Additionally, cross-pipeline consistency checks are carried out by comparing detections from multiple search pipelines (e.g., GstLAL, PyCBC, MBTA, etc) to verify agreement in classification and parameter recovery. Together, these measures provide a robust mechanism for validating \ac{LLAI} performance and ensuring that any updates or new features integrated through \ac{MEG} maintain the expected reliability and accuracy standards.

\begin{figure*}[ht!]
\centering
\includegraphics[width=0.5\linewidth]{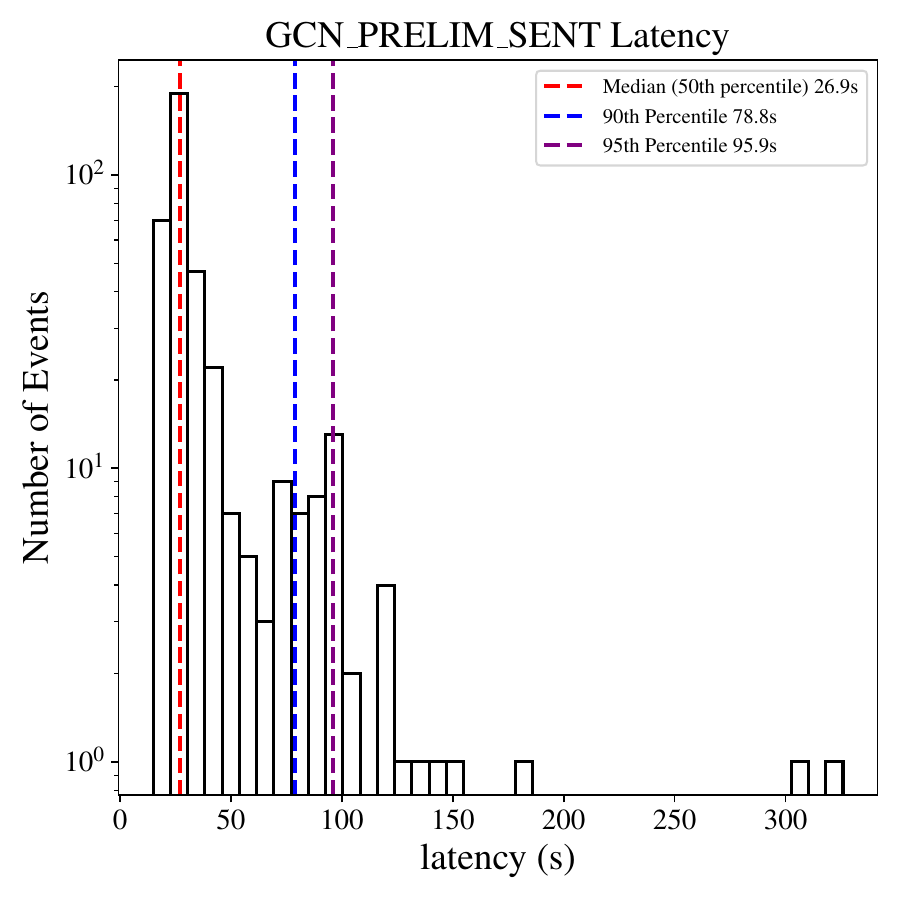}

\caption{Histogram of latencies for the sending of the GCN preliminary alert in cycle 17. In this cycle, triggers from GstLAL, MBTA, PyCBC, SPIIR, cWB, and MLy were replayed using MEG.}
\label{fig:meg_latency}
\end{figure*}

The inclusion of \ac{MEG} in the latest \ac{MDC} (see Sec.~\ref{LLPIC}) framework provides confidence that the low-latency alerts distributed to the astronomical community are scientifically robust. The replay-and-review framework has enabled verification of new data products — such as sub-solar mass classifications and binned chirp mass estimates -- and a new \ac{ML} based search pipeline Aframe \cite{Marx:2024wjt} prior to their release on \ac{GraceDB}. Figure \ref{fig:meg_latency} shows the \ac{GCN} preliminary alert latencies for cycle 17 of \ac{LLPIC}. During this cycle, \ac{MEG} was used to replay triggers for GstLAL, PyCBC, MBTA, SPIIR, cWB and MLy pipelines whereas Aframe analyzed the real-time data to produce \ac{GW} triggers. \ac{MEG} was successfully able to achieve a median latency similar to that of cycle 4 where individual pipelines were analyzing the real-time data and producing triggers.  

As the LVK moves toward O5, we plan to extend the LLPIC framework's validation capabilities, using MEG to create replay events and reproduce their associated data products. This includes extending the validation workflow to cover low-latency Bayesian classification \citep{Romero-Shaw:2020owr} and potentially machine-learning–based event vetting. Another priority is to automate more of the validation workflow through standardized scripts, reducing the need for manual inspection during daily replay checks. The long-term goal is to support continuous integration–style testing within \ac{LLPIC}, with MEG providing replay events and the validation workflow assessing downstream low-latency features before deployment.

\clearpage

\bibliography{references}{}
\bibliographystyle{aasjournal}

\end{document}